\documentclass[journal]{IEEEtran}
\usepackage{cite}
\usepackage{amsmath,amssymb,amsfonts}
\usepackage{algorithm}               %format of the algorithm
\usepackage{algorithmic}             %format of the algorithm
\usepackage{graphicx}
\usepackage{textcomp}
\usepackage{xcolor}
\usepackage{setspace}
\usepackage{subfigure}
\usepackage[caption=false,font=footnotesize]{subfig}
\usepackage{booktabs}
\usepackage{multirow} 
\usepackage{diagbox}
\usepackage{color}
\usepackage{stfloats}
\newcommand{\R}[1]{\textcolor{black}{#1}}
\begin{document}
	
		\bstctlcite{IEEEexample:BSTcontrol}
	\title{ 
		Unified Timing Analysis for Closed-Loop Goal-Oriented Wireless Communication
		
		\thanks{
			The work of Lintao Li and Wei Chen was supported by the National Natural Science Foundation of China/Research Grants Council Collaborative Research Scheme under Grant No. 62261160390. The work of A. E. Kalør was supported in part by the Independent Research Fund Denmark (IRFD) under Grant 1056-00006B and in part by the Horizon Europe SNS project “6G-GOALS” (grant 101139232).
			The work of P. Popovski was supported in part by the Villum Investigator Grant “WATER” from the Velux Foundation, Denmark, and in part by the Horizon Europe SNS project “6G-GOALS” (grant 101139232).

			Lintao Li and Wei Chen are with the Department of Electronic Engineering, Tsinghua University, Beijing 100084, China, and also with the Beijing National Research Center for Information Science and Technology, Tsinghua University, Beijing 100084, China (email: llt20@mails.tsinghua.edu.cn; wchen@tsinghua.edu.cn).

      Anders E. Kalør is with the Department of Information and Computer Science, Keio University, Yokohama 223-8522, Japan, and with the Department of Electronic Systems, Aalborg University, 9220 Aalborg, Denmark (e-mail: aek@keio.jp).
			
			Petar Popovski is with the Department of Electronic Systems, Aalborg University, 9220 Aalborg, Denmark (e-mail: petarp@es.aau.dk).}
	}

	\author{Lintao Li,  Anders E. Kalør,~\IEEEmembership{Member,~IEEE,} Petar Popovski,~\IEEEmembership{Fellow,~IEEE}, and Wei Chen,~\IEEEmembership{Senior Member,~IEEE}    }
	
	\maketitle
	
	\begin{abstract}
		Goal-oriented communication has become one of the focal concepts in sixth-generation communication systems owing to its potential to provide intelligent, immersive, and real-time mobile services. The emerging paradigms of goal-oriented communication constitute closed loops integrating communication, computation, and sensing. However, challenges arise for closed-loop timing analysis due to multiple random factors that affect the communication/computation latency, as well as the heterogeneity of feedback mechanisms across multi-modal sensing data. To tackle these problems, we aim to provide a unified timing analysis framework for closed-loop goal-oriented communication (CGC) systems over fading channels. The proposed framework is unified as it considers computation, compression, and communication latency in the loop with different configurations. To capture the heterogeneity across multi-modal feedback, we categorize the sensory data into the periodic-feedback and event-triggered, respectively. We formulate timing constraints based on average and tail performance, covering timeliness, jitter, and reliability of CGC systems. A method based on saddlepoint approximation is proposed to obtain the distribution of closed-loop latency. The results show that the modified saddlepoint approximation is capable of accurately characterizing the latency distribution of the loop with analytically tractable expressions. This sets the basis for low-complexity co-design of communication and computation.
	\end{abstract}
	

	\section{Introduction}
	
	\subsection{Motivation and Contributions}
	Driven by the rapid development of information technology and the continuous improvement of hardware capability, the sixth-generation (6G) communication systems are envisioned to fuse the physical and digital worlds by providing intelligent, immersive, and real-time services~\cite{popovski2022}. This is, for instance, reflected in the IMT-2030 draft for the 6G wireless communication technology \cite{itu} published by the International Telecommunication Union (ITU), which in addition to hyper reliable and low-latency communication also includes immersive communication as central 6G usage scenarios.
	Against this background, the goal-oriented communication paradigm gets significant traction in 6G~\cite{gunduz2023}. 
	Supported by artificial intelligence (AI), goal-oriented communication aims to provide intelligent and immersive experiences for users by optimizing the communication for the receiver's intention and experience, rather than simply providing reliable, content-blind packet delivery~\cite{gunduz202310,6ggoals}. Combined with Metaverse and Digital Twin (DT) technology, this enables 6G to break the spatial constraints to promote the achievement of the flexible interconnection between the physical world and the digital virtual space~\cite{xu2023}.

	Differently from 5G, goal-oriented paradigms require demanding \emph{closed-loop} interactions with strict timing requirements, starting from the commands transmitted by the user and ending with the reception of feedback. The closed-loop interactions are common in several of the typical 6G scenarios considered by the Third Generation Partnership Project (3GPP) \cite{3gpp}, including remote control, virtual social, and gaming. However, the closed-loop timing analysis is challenging compared to the point-to-point perspective that has dominated the literature so far, as it needs to consider processing latency of intermediate notes caused by, e.g., compression and computation. Thus, there is a need to develop a unified timing-analysis framework for closed-loop goal-oriented communication (CGC) systems over wireless channels, which can serve as a tool to ensure that users get the desired real-time experience.
	
\begin{figure*}[t]
		\centerline{\includegraphics[width=17.5cm]{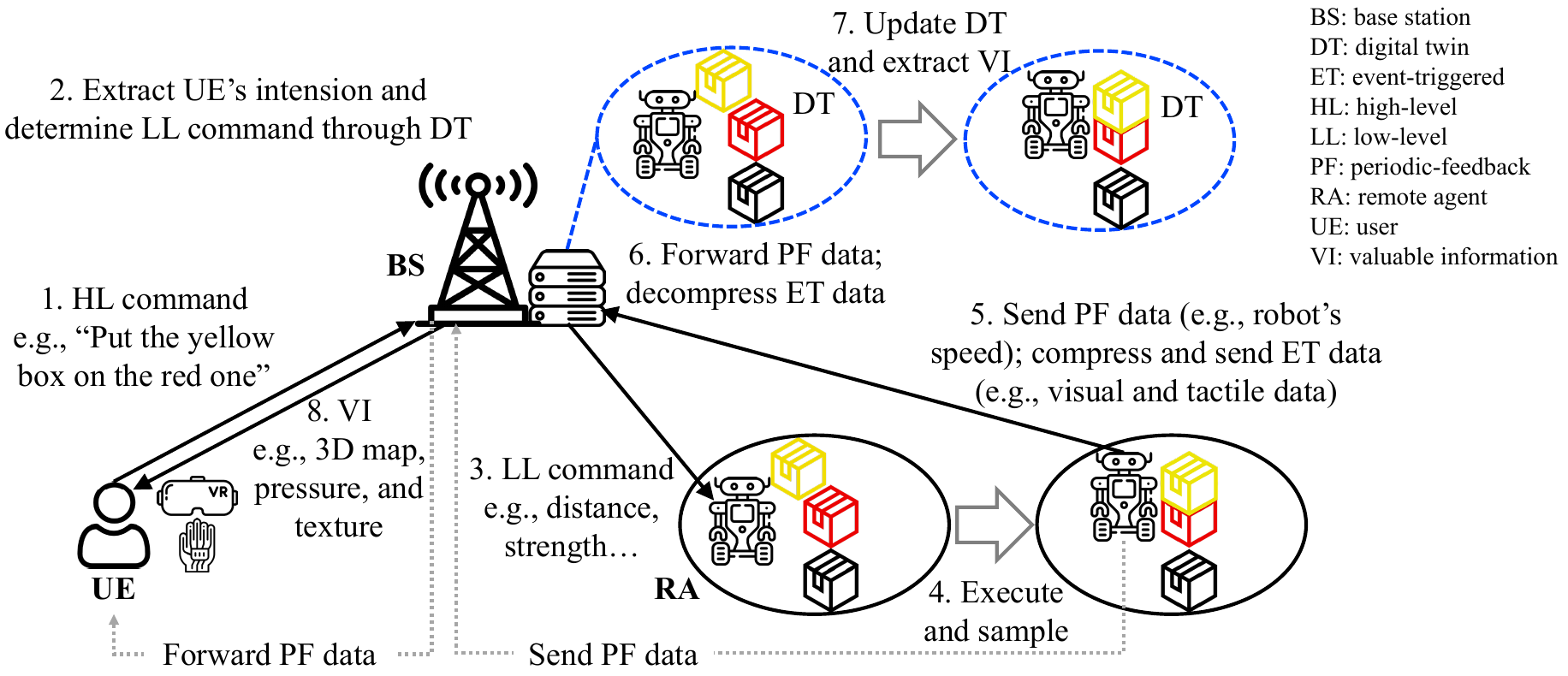}}
		\caption{System model for a general CGC system. This figure shows an example of the remote control with the assistance of MEC, semantic information, and DT. Multiple communication and computation processes are included in this closed loop.}
		\label{sysmod}
	\end{figure*}

	To address this issue, we aim to formulate a unified framework for timing analysis in CGC systems, accounting for communication, computation, and compression latency. Specifically, we consider a general CGC system, comprising a user equipment (UE), a base station (BS), and a remote agent (RA), all assumed to operate in a closed loop. The system is exemplified as a robotic scenario with a DT at the BS in Fig. \ref{sysmod}. We consider both periodic and sporadic traffic. The periodic traffic, referred to as periodic feedback (PF), is transmitted at fixed intervals from the RA to UE, representing, e.g., sensor data. The sporadic traffic is initiated by control data sent from the UE to RA, which then responds with event-triggered (ET) data. Communication occurs though the BS, which is equipped with a mobile-edge-computing (MEC) server to handle computation tasks, such as feature extraction, compression, and decompression. Although the remote agent is depicted as a robot, it could also function as a cloud server in a gaming scenario or as another UE in a virtual social scenario. A key novelty of the proposed model is its capability to enable closed-loop analysis of heterogeneous multi-modal feedback, comprising both event-triggered and periodic feedback, along with stochastic computation and compression latency. \R{In goal-oriented systems with various goals, such as metaverse systems that ensure an immersive experience for UEs, these elements are generally present.} The stochastic nature of the computation and compression latency has been ignored in most existing work, although they are crucial for end-to-end latency analysis~\cite{suman2023,han2019,lorch2001}. Besides, motivated by perceptual requirements of interactive applications~\cite{3gpp,Steinbach2012,antonakoglou2018,petar2024arxiv}, we explicitly model the asynchrony between the transmission latency of PF and ET data, referred to as \emph{jitter}. Further details of the model will be given in Section II.

	Note that the model in Fig.~\ref{sysmod} is rather general, capable to cover some commonly referred scenarios for CGC systems. For the remote control task shown in Fig.~\ref{sysmod}, we are interested in the closed-loop latency by accounting for the timeliness of the control data and the jitter of the multi-modal feedback. For the authorization of avatar usage rights, specified by 3GPP in \cite{3gpp}, the RA acts as a cloud server. UE sends requests to the cloud server through BS without relying on the computation process at the BS. Once the cloud server receives the requests, it inquires and returns profiles of the avatar to the UE. Meanwhile, authorization information will be sent to UE periodically to authorize the use of the avatar legally. In this scenario, reducing the asynchrony between profile reception and authorization ensures immediate avatar utilization without additional latency. Differently from the remote control, we focus on the closed-loop latency without a specific concern of control link latency.

	Using the proposed model, and with consideration of timeliness, jitter, and reliability, we formulate a unified framework of timing analysis for CGC systems,capable of addressing both average- and tail-based timing constraints for CGC systems with different information-processing mechanisms and configurations. 
	To this end, we propose a general saddlepoint-approximation-based method to obtain the analytical expression of the distribution of closed-loop latency with high accuracy. \R{Then a low-complexity joint SPA and CLT-based approximation method is proposed. Based on the  joint SPA and CLT-based approximation method,} we also give a corollary for efficiently approximating the conditional expectation given the timing constraints. The proposed methods can be generalized to timing analysis in other closed-loop scenarios.

	The main contributions are summarized as follows: 
	\begin{itemize}
		\item We formulate a unified timing-analysis framework for CGC systems with different configurations, which takes the randomness of the computation, compression, and communication latency into account. The formulated framework integrates the timeliness, jitter, and reliability of CGC systems.
		\item We consider two heterogeneous information flows with both PF and ET data, representing different feedback mechanisms of multi-modal sensing data. 
		\item We propose a general saddlepoint approximation (SPA)-based method for analytically characterizing the timing measures in CGC systems. For specific scenarios,  the central limit theorem (CLT) is also \R{utilized} to improve the efficiency of the method while ensuring accuracy. 
		\item \R{Based on the joint SPA and CLT-based approximation, an efficient approximation is presented for calculating the conditional expectation given timing constraints.}
	\end{itemize}

	\begin{table*}[h]
	\small 
	\centering 
	\caption{\R{Comparison with related works}} 
	\begin{center} 
		\R{ 
			\begin{tabular}{ccccc} 
				\toprule  
				\textbf{Work} &  \textbf{Computation Time} & \textbf{Feedback Mechanism} & \textbf{Analyzing Method}  & \textbf{Considered Scenarios}  \\\midrule
				\cite{Cao2023} &  Not consider & Event-triggered & Queueing theory & Remote control \\
				\cite{li2021}  & Deterministic & Event-triggered & SPA for i.i.d. random variables & Federated learning\\
				\cite{suman2023} & Stochastic & Event-triggered & Truncated convolution& Remote control \\
				This work & Stochastic & Event-triggered, periodic & SPA for non-i.i.d. random variables & General CGC systems\\ \bottomrule
		\end{tabular}  }
	\end{center}  
\end{table*}

\begin{table}[t]
		\small 
		\centering 
		\caption{\R{Acronyms used in the paper} }
  \vspace{-4mm}
		\R{\begin{center}  
				\begin{tabular}{cp{60mm}} 
					\toprule  
					\textbf{Abbreviation} & \textbf{Definition} \\ \midrule 
					CGC & closed-loop goal-oriented communication  \\
					DT & digital twin \\
					AI & artificial intelligence \\
					UE & user equipment \\
					BS & base station \\
					RA & remote agent \\
					PF & periodic feedback \\
					ET & event-triggered \\
					MEC& mobile edge computing \\
					CLT& central limit theorem \\
					SPA& saddlepoint approximation \\
					CDF & cumulative distribution function \\ 
					PDF & probability density function \\
					PMF& probability mass function \\
					FL &  feedback link \\
					CL & control link \\ 
					VI & valuable information \\
					FSD & frequency or spatial diversity \\
					HL & high-level \\
					LL & low-level \\
					CD & compressed data  \\
					SNR &  signal-to-noise ratio \\
					CSIRT  & channel state information known \\
					& at both receiver and transmitter \\ \bottomrule
				\end{tabular}  
		\end{center} }
	\end{table}

	\subsection{Related Work}
	
	In \cite{Kizilkaya2023}, a goal-oriented communication and prediction co-design was proposed to improve the reliability of the haptic communication system. The end-to-end communication latency was chosen as the timing metric with a queueing model, while the computation latency is omitted in \cite{Kizilkaya2023}. Wen \emph{et al.} optimized the discriminant gain to achieve the goal of improving the accuracy of AI inference in an integrated sensing and communication system \cite{Wen2023}. Meanwhile, they considered the latency constraint in this system, which focused on the uplink latency comprised of communication, computation, and sensing procedures. The authors of \cite{van2022} designed resource allocation and computation offloading schemes to minimize the uplink latency of DT-enabled Metaverse. In \cite{Wen2023} and \cite{van2022}, the constant computation latency is incorporated into the timing measurements. However, they only considered the one-way latency instead of the closed-loop latency.

	Conducting closed-loop timing analysis was proved to be effective for the wireless networked control system in \cite{aol} and \cite{Cao2023}. Thus, it is important to analyze the timing performance of a goal-oriented communication system from the closed-loop perspective.  In \cite{li2021}, the authors carried out the closed-loop timing analysis for wireless federated learning systems by taking both random latency of local information uploading and global model broadcasting into account. However, the computation latency was also assumed as a deterministic term in \cite{li2021}, which ignored its stochastic property \cite{suman2023,han2019,lorch2001}. To address these problems, Suman \emph{et al.} conducted a statistical timing analysis for a closed-loop teleoperation system \cite{suman2023}, in which the randomness of computation and compression latency were accounted for. However, the heterogeneity between the multi-modal feedback, which widely exists in various goal-oriented communication, was not considered.

	The SPA technique that we apply is a powerful method to approximate complicated distributions 
	with high accuracy guarantees while avoiding intractable
	computations \cite{LR}. SPA has been widely used in communication research. In \cite{li2021}, SPA was used to characterize the distribution of transmission latency in wireless federated learning systems. An efficient SPA-based method for evaluating block error probabilities was proposed in \cite{kislal2023}, and the power-latency-throughput tradeoff of delay-constrained wireless systems was characterized by SPA in \cite{icc1}.

	\subsection{Paper Organization and Notation}
	
	The paper is structured as follows. Section II presents the unified timing framework for CGC systems considered in this paper. Section III introduces the latency components and formally defines the overall objective, while Section IV proposes the SPA-based timing analysis for the CGC system. Section V validates the analysis through numerical results, and finally the paper is concluded in Section VI.
	
	\emph{Notation:} $\mathbb{E}\{x\}$ denotes the expectation of the random variable $x$, and ${\rm Var}\{x\}$ denotes its variance. $f'(x)$ and $f''(x)$ denote the first-order and second-order derivative of $f(x)$, respectively. $\mathcal{N}(0,1)$ denotes the standard Gaussian distribution with cumulative distribution function (CDF) $\Phi(x)=\frac{1}{\sqrt{2\pi}}\int_{-\infty}^{x}e^{-\frac{y^2}{2}}{\rm d}y$ and probability density function (PDF) $\phi(x)=\frac{1}{\sqrt{2\pi}}e^{-\frac{x^2}{2}}$. $\gamma(s,x)=\int_{0}^{x}t^{s-1}e^{-t}{\rm d}t$ is the lower incomplete gamma function, $\{x\}^+=\max\{x,0\}$, and ${\rm sign}(x)$ denotes the sign of $x$.

	\section{A Unified Framework for Timing Analysis in CGC Systems with Different Configurations}
	
	In this section, we present a unified framework for timing analysis in CGC systems. \R{We begin by providing an overview of the system, followed by an introduction to latency components. Section II-C outlines the wireless channel model, while Section II-D specifies different timing constraints employed to formulate the proposed framework with diverse configurations.}

	\subsection{System Overview}

\begin{figure}[t]
		\centerline{\includegraphics[width=8.5cm]{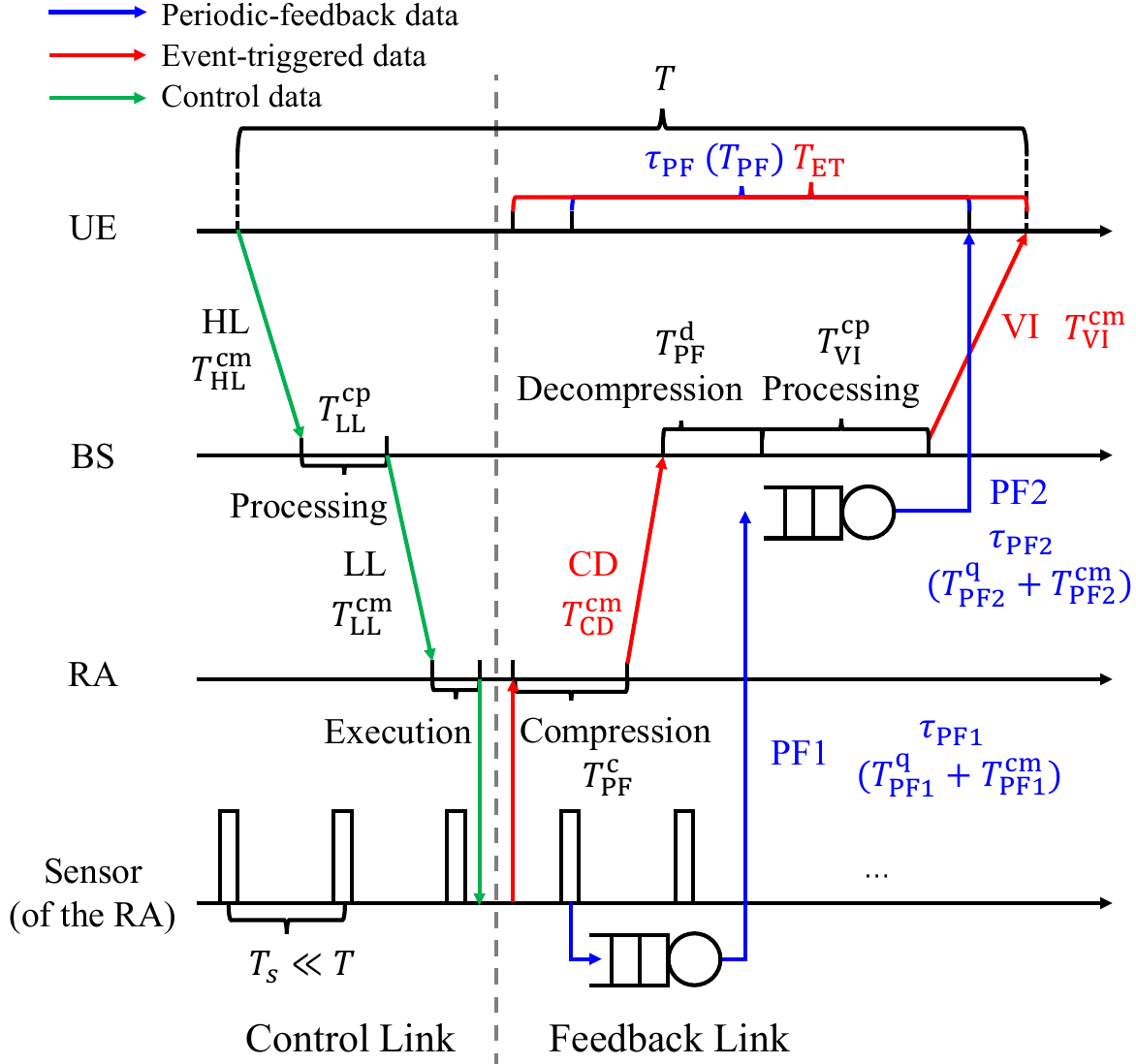}}
		\caption{Diagram of the components of the closed-loop latency. The UE transmits commands to the RA through the control link, and the RA transmits both periodic-feedback and event-triggered sensor data to the UE.}
		\label{bsysmod}
	\end{figure}

	\R{We study the CGC system illustrated} in Fig.~\ref{sysmod}, which comprises a BS, a UE, and a RA equipped with a sensor. \R{The UE and the RA communicate over a wireless channel via the BS, using dedicated channel resources.} Additionally, the BS is equipped with an MEC server to perform computation tasks, \R{including feature extraction and decompression.}  \R{We designate the channel from the UE to the RA as the \emph{control link} (CL), while the link from the RA to the UE is the \emph{feedback link} (FL).}  The communication flow is depicted in Fig. \ref{bsysmod}, where we consider \R{three types of data: control data}, periodic-feedback (PF) data, and event-triggered (ET) data.
	
	 Control data is transmitted sporadically by the UE as high-level (HL) commands,  \R{assumed to be sent at relatively long intervals to eliminate the need for queuing. For instance, control data may represent the UE's intention to place the yellow box on the red one, as shown in Fig. \ref{sysmod}. The HL commands are subject to a random processing delay at the BS.} Specifically, the UE transmits an HL command to the BS, which then processes the received command to extract the semantic information, \R{for instance, with the assistance of} a large language model \cite{llm} and the DT model maintained at the BS. The HL command is assumed to consist of $N_{\rm HL}$ packets, each of size $n_{\rm HL}$ bits each\footnote{\R{Typically, there is one packet containing the specific command signals. However, the analysis in this paper does not specify the value of the number of packets for both HL and low-level, as the proposed framework and analysis are general.}}. At the BS, the HL command is \R{converted} into detailed control signals, referred to as low-level (LL) commands, which are transmitted from the BS to the RA. The LL commands consist of $N_{\rm LL}$ packets, \R{each containing $n_{\rm LL}$ bits.}

	  \R{PF data is typically small in size and} represents numeric values, such as the speed of the RA and the temperature of the remote environment \cite{3gpp}. The PF data is collected at fixed, short intervals every $T_{\rm s}$ seconds and can be \R{displayed} on UE without further processing. However, the PF data is queued at both the RA and the BS. \R{We denote $n_{\rm PF}$ (in bits) as the size of a PF packet.}
    
    Finally, ET data is generated by the RA \R{in response to} an LL command. \R{ET data is generally throughput-intensive and may include multimedia data such as video, tactile details, and audio. Thus, the RA first compresses the ET data before transmitting it to the BS.} The MEC server at BS decompresses the data, extracts valuable information (VI) from the ET data, such as the location and texture of the objective, and updates its DT model \R{in preparation for} the next loop. After processing, the VI is encoded into $N_{\rm VI}$ packets, \R{each with an equal size of $n_{\rm VI}$ bits.The VI is then transmitted to the UE.
    The total size of the original ET data at the RA is denoted as $n_{\rm ET}$ (in bits). The compressed data (CD) is transmitted using $N_{\rm CD}$ packets, each containing $n_{\rm CD}$ bits.}

	The focus of this paper is on the closed-loop latency, $T$, of a single transmission round under various constraints on the individual links. Specifically, a single transmission round \R{begins with} the transmission of an HL command by the UE and lasts until the UE has received both the corresponding ET data \emph{and} the PF data generated closest to the time instant \R{when} the LL data is received by the RA. The proposed framework in the following subsections \R{incorporates} the latency requirements for CL and the jitter between two heterogeneous information flows in FL. These requirements will affect $T$ by influencing the latency of CL and FL. \R{In scenarios involving multiple transmissions, additional queueing delay is introduced.} We leave the analysis of the scenario with multiple transmissions for future work.

	\subsection{Latency Components}
	Based on the previous description and Fig. \ref{bsysmod}, the transmission latency of the PF data, denoted by $T_{\rm PF}$, is given by
	\begin{equation}
		\begin{aligned}
			T_{\rm PF}=
			T^{\rm q}_{\rm PF1}+T^{\rm cm}_{\rm PF1}+T^{\rm q}_{\rm PF2}+T^{\rm cm}_{\rm PF2},
		\end{aligned}
	\end{equation}
	 where $T^{\rm q}_{\rm PF1}$ and $T^{\rm cm}_{\rm PF1}$ \R{denote} the queueing and transmission delays from the RA to the BS, respectively, and $T^{\rm q}_{\rm PF2}$ and $T^{\rm cm}_{\rm PF2}$ \R{denote} the queueing and transmission delays from the BS to the UE, respectively. For the control data, the latency from UE to RA can be written
	\begin{equation}
		T_{\rm CL}=T^{\rm cm}_{\rm HL}+T_{\rm LL}^{\rm cp}+T^{\rm cm}_{\rm LL},
	\end{equation}
	where $T^{\rm cm}_{\rm HL}$ and $T^{\rm cm}_{\rm LL}$ \R{denote} the transmission delay of the HL command and the LL command, respectively, and $T_{\rm LL}^{\rm cp}$ \R{denotes} the processing delay at the BS. Similarly, \R{the latency of transmitting ET data} from the RA to the UE is
	\begin{equation}\label{timemd}
		T_{\rm ET}=T^{\rm c}_{\rm ET}+T^{\rm cm}_{\rm CD}+T^{\rm d}_{\rm ET}+T^{\rm cp}_{\rm VI}+T^{\rm cm}_{\rm VI},
	\end{equation}
	where $T^{\rm c}_{\rm ET}$ \R{denotes} the compression delay at the RA, $T^{\rm cm}_{\rm CD}$ and $T^{\rm cm}_{\rm VI}$ \R{denote} the transmission delay from the RA to the BS and from the BS to the UE, respectively, and $T^{\rm d}_{\rm ET}$ and $T^{\rm cp}_{\rm VI}$ \R{denote} the decompression delay and the processing delay at the BS, respectively.
	
	Recall that we are interested in the closed-loop latency from the generation of the HL command until the reception of both the corresponding ET data and the PT data generated closest in time to the generation time of the ET data. \R{Given that} the interval of the periodic transmission, $T_{\rm s}$, is relatively small, its impact is negligible, and the latency of the feedback link, denoted as $T_{\rm FL}$, can thus be approximated by
	\begin{equation}
		T_{\rm FL}\approx\max\left\{T_{\rm PF},T_{\rm ET}\right\}.
	\end{equation}
	Using this, we finally write the closed-loop latency as
	\begin{equation}
		T=T_{\rm CL}+T_{\rm FL}.
	\end{equation}
	
	Finally, we also define the jitter of the feedback link as the absolute difference between $T_{\rm ET}$ and $T_{\rm PF}$, i.e., $|T_{\rm ET}-T_{\rm PF}|$.

	\subsection{Wireless Channel}

	We assume independent block fading channels at all links and the channel gain remains constant over the length of a transmitting packet. Assuming a transmission power of $P_i$, at link $i \in \{{\rm HL},{\rm LL},{\rm PF1}, {\rm PF2},{\rm CD},{\rm VI}\}$, the signal-to-noise ratio (SNR) at the receiver is given by
	\begin{equation}
		\!\!\! {\rm SNR}_{i}=\frac{d_i^{-\ell_i}|h_i|^2P_i}{N_0B_i},\, \text{$i \in \{{\rm HL},{\rm LL},{\rm PF1}, {\rm PF2},{\rm CD},{\rm VI}\}$},
	\end{equation}
	where $|h_i|^2$ is the small-scale channel power gain, $d_i$ is the distance between the transmitter and the receiver, $\ell_i$ is the path-loss exponent, $N_0$ is the spectral density of the noise power, and $B_i$ is the bandwidth.  For $i\in \{{\rm LL,CD, PF1}\}$, $d_i$ refers to the distance between the RA and BS, which we denote \R{as} $d_{\rm RB}$. For $i=\{\rm HL,PF2,VI\}$, $d_{i}$ refers to the distance between the BS and UE, which we denote \R{as} $d_{\rm BU}$. Besides, we let $\ell_{\rm BU}$ denote the path-loss exponent of the link between the BS and UE, while $\ell_{\rm RB}$ \R{denotes} the path-loss exponent of the link between the RA and BS. Moreover, $B_{\rm PF1}=B_{\rm PF2}=B_{\rm PF}$, while $B_{\rm CD}=B_{\rm VI}=B_{\rm ET}$. \R{In the case where channel state information is known only at the receiver side (CSIR)}, the transmitter adopts the $\epsilon_i$-outage rate $R_i$. With $|h_i|^2\sim {\rm exp}(1)$, the relationship between $R_i$ and $\epsilon_i$ is given by \cite{oc}
	\begin{equation}\label{orate}
		R_i=B_i\log_2\left(1-\frac{d_i^{-\ell_i}P_i}{N_0B_i}\ln(1-\epsilon_i)\right). 
	\end{equation}
	For completeness, we also consider the case in which the transmitter also has perfect channel state information for the transmission of PF packets, referred to as CSIRT. In this case, we \R{further} assume that frequency or spatial diversity (FSD) is available, e.g., through multiple antennas. This is motivated by our previous work \cite{twc1,icc2024}, which \R{demonstrates} that with both CSIRT and FSD, \R{the delay violation probability can be ensured to be zero with finite average power.} In this case, we let $\tau_{\rm PF}$ denote the deadline \R{for} PF data with CSIRT and FSD, \R{indicating that} the latency of each PF packet \R{can not exceed} $\tau_{\rm PF}$. Thus, with CSIRT and FSD, we can use $\tau_{\rm PF}$ to represent the latency of PF data.\footnote{\R{Although} the average power is finite, the required instantaneous power may be large. If there is a maximum power constraint, the delay violation probability of the packet is not zero and can be characterized by large deviation theory \cite{twc1} and extreme value theory \cite{li2021}. The analysis of this setting is left for future work.} Moreover, we let $\tau_{\rm PF1}$ and $\tau_{\rm PF2}$ denote the deadlines for the PF data transmission from the RA to BS and the BS to UE, respectively, \R{satisfying} $\tau_{\rm PF1}+\tau_{\rm PF2}\leq \tau_{\rm PF}$.
	
	\begin{figure}[t]
		\centerline{\includegraphics[width=8.5cm]{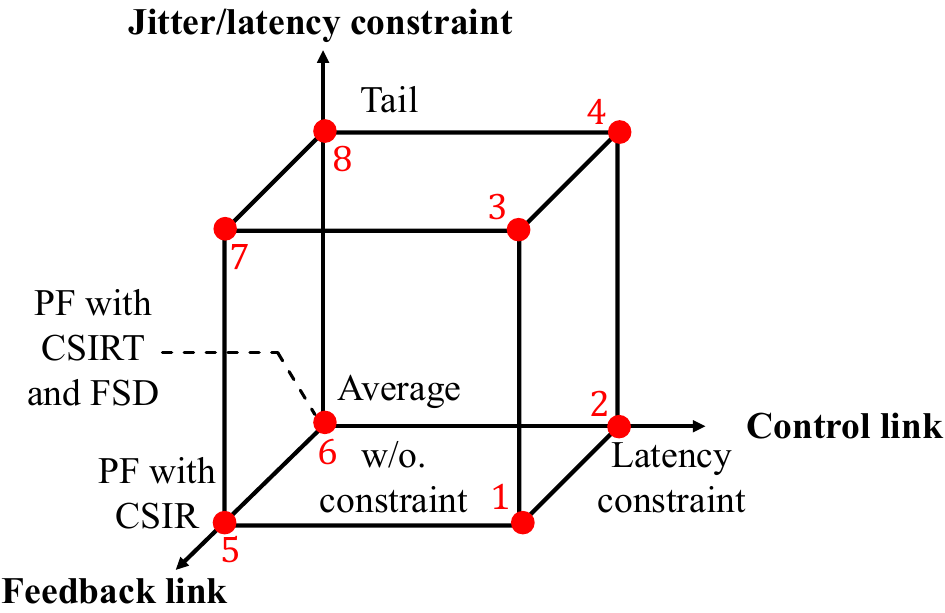}}
		\caption{Configurations considered in the proposed framework.}
		\label{constraints}
	\end{figure}

	\subsection{A Unified Framework of Timing Constraints for CGC Systems with Different Configurations}
	
	In this subsection, we establish a unified timing-analysis framework for CGC systems under different configurations, shown as vertices in Fig.  \ref{constraints}. The proposed framework ensures the timeliness performance of CL and the jitter performance of FL to cover general requirements for CGC systems. These constraints also influence the closed-loop latency $T$ through $T_{\rm CL}$ and $T_{\rm FL}$ to ensure the reliability of CGC systems.

	As \R{depicted} on the jitter/latency constraint axis, we consider both average and tail-based constraints on jitter and latency. \R{Average-based constraints} are related to the expectation of timing metrics, while tail-based constraints are related to the violation probability of timing metrics. For the CL, which handles the transmission of control data, we \R{account for both configurations} with and without a latency constraint representing a timeliness requirement on $T_{\rm CL}$. \R{Additionally}, the FL may or may not have CSIRT and FSD for the PF data, as represented by the last axis. \R{When CSIRT and FSD are available, a hard delay constraint is imposed on the PF data, denoted by $\tau_{\rm PF}$, as discussed in Section II-C. Under this condition, the PF transmission latency $T_{\rm PF}$ is treated as deterministic, where $T_{\rm PF}=\tau_{\rm PF}$. When only CSIR is available, $T_{\rm PF}$ becomes a random variable.}
	
	 \R{Regarding delay jitter, two different evaluation methods correspond to the different configurations along the FL axis in Fig. \ref{constraints}. The first method applies when the transmission of PF data benefits from both CSIRT and FSD.} In this case, the latency of PF data is \R{determined} by the threshold $\tau_{\rm PF}$.  \R{Consequently}, the jitter is defined as $|T_{\rm ET}-\tau_{\rm PF}|$.  
  
	\R{The second method applies when the transmission of PF data only has CSIR, resulting in the stochastic latency of transmitting PF data.} \R{In this case}, jitter is the difference between two random variables, \R{expressed as $|T_{\rm ET}-T_{\rm PF}|$.}
	
	Based on these constraints \R{along} each axis, we can combine them to \R{generate} the eight vertices of the cube \R{illustrated} in Fig. \ref{constraints}.\footnote{Note that except the constraints \R{at} the vertices of the cube, \R{intermediate constraints can also be formed, such as a combination of tail-based FL constraints with average-based CL constraints. These constraints can be visualized as points between vertices $i$ and $i+2$, where$i\in\{1,2,5,6\}$.} Due to space limits, we omit detailed discussions.}  In the following, we elaborate on each vertex in Fig. \ref{constraints}. Vertices 1-4 refer to interactive systems where RA is controlled under different configurations as shown in Fig. \ref{sysmod}.
	\begin{itemize}
		\item Vertex 1 -- average constraints for CL and PF with CSIR: 
		\begin{equation}\label{v1}
			\begin{aligned}
				\begin{cases}
					\mathbb{E}\left\{T_{\rm CL}\right\}\leq \varrho_{\rm CL},\\
					\mathbb{E}\left\{\left|T_{\rm ET}-T_{\rm PF}\right|\right\}\leq \varrho_{\rm FLd} ,
				\end{cases}
			\end{aligned}
		\end{equation}
		where $\varrho_{\rm CL}$ \R{represents} the average latency constraint of CL, and $\varrho_{\rm FLd}$ \R{denotes} the average jitter constraint of FL.

		\item Vertex 2 -- average constraints for CL and PF with CSIRT and FSD:
		\begin{equation}\label{asf}
			\begin{aligned}
				\begin{cases}
					\mathbb{E}\left\{T_{\rm CL}\right\}\leq \varrho_{\rm CL},\\
					\tau_{\rm PF1}+\tau_{\rm PF2}\leq \tau_{\rm PF}, \\
					\mathbb{E}\left\{\left(T_{\rm ET}-\tau_{\rm PF}\right)^+\right\}\leq \varrho_{\rm FLs} ,
				\end{cases}
			\end{aligned}
		\end{equation}
		where  $\tau_{\rm PF}$ and $\varrho_{\rm FLs}$ are average jitter constraints of FL. 
  In this case, we treat the part in which $T_{\rm ET}<\tau_{\rm PF}$ as no jitter. This is because $\tau_{\rm PF}$ is deterministic so that both ET and PF feedback can be shown to UE until PF data is received if $T_{\rm ET}<\tau_{\rm PF}$. Note that the third constraint in this case can be rewritten as
		\begin{equation}
			\mathbb{E}\left\{T_{\rm ET}-\tau_{\rm PF}|T_{\rm ET}>\tau_{\rm PF}\right\}\leq\frac{\varrho_{\rm FLs}}{	\Pr\{T_{\rm ET}>\tau_{\rm PF}\}},
		\end{equation}
		which can be seen as a condition value-at-risk (CVaR) based constraint \cite{bennis2018}.
		
		\item Vertex 3 -- tail constraints for CL and PF with CSIR:
		\begin{equation}\label{v3}
			\begin{aligned}
				\begin{cases}
					\Pr\left\{T_{\rm CL}<\tau_{\rm CL}\right\}\geq \eta_{\rm CL},\\
					\Pr\left\{|T_{\rm ET}-T_{\rm PF}|<\tau_{\rm td}\right\}\geq \eta_{\rm td},\\
					{\rm Var}\left\{|T_{\rm ET}-T_{\rm PF}|\right\}\leq \rho_{\rm td},
				\end{cases}
			\end{aligned}
		\end{equation}
		where $\eta_{\rm CL}$ and $\tau_{\rm CL}$ \R{represent the} tail latency constraints of CL. \R{The parameters} $\tau_{\rm td}$, $\eta_{\rm td}$, and $\rho_{\rm td}$ \R{denote} the tail jitter constraints of FL. For the first constraint in Eq. \eqref{v3}, if $\eta_{\rm CL}$ is \R{specified,  the minimum $\tau_{\rm CL}$ becomes the} value at risk (VaR) \cite{bennis2018}, i.e.,
		\begin{equation}
			\tau_{\rm CL}=\inf_{t}\left\{\Pr\{T_{\rm CL}<t\}\geq\eta_{\rm CL}\right\}.
		\end{equation}
		If $\tau_{\rm CL}$ is given, solving the maximum value of  $\eta_{\rm CL}$ is equal to minimizing the delay violation probability, i.e.,
		\begin{equation}
			\eta_{\rm CL}=\inf_{p}\left\{\Pr\{T_{\rm CL}\geq \tau_{\rm CL}\}\leq p\right\}.
		\end{equation}
		
		The second and third constraints in Eq. \eqref{v3} refer to the tail jitter constraint for FL. \R{In addition to} the violation probability of $|T_{\rm ET}-T_{\rm PF}|$, the variance of the jitter is included to cover general cases, \R{serving as} an important measurement for tail behavior, \R{as demonstrated} in \cite{Liu2019}.

		\item Vertex 4 -- tail constraints for CL and PF with CSIRT and FSD:
		\begin{equation}\label{sdf}
			\begin{aligned}
				\begin{cases}
					\Pr\left\{T_{\rm CL}<\tau_{\rm CL}\right\}\geq \eta_{\rm CL},\\
					\tau_{\rm PF1}+\tau_{\rm PF2}\leq \tau_{\rm PF}, \\
					\Pr\{T_{\rm ET}<\tau_{\rm PF}\}\geq \eta_{\rm ts},\\
					{\rm Var}\{T_{\rm ET}-\tau_{\rm PF}\}\leq \rho_{\rm ts},
				\end{cases}
			\end{aligned}
		\end{equation}
	where $\tau_{\rm PF}$, $\eta_{\rm ts}$ and $\rho_{\rm ts}$ \R{represent} the tail jitter constraints of FL. Since $\tau_{\rm PF}$ is a constant, the fourth constraint in Eq. \eqref{sdf} is equivalent to 
		\begin{equation}\label{equicon}
			{\rm Var}\left\{T_{\rm ET}\right\}\leq \rho_{\rm ts}.
		\end{equation}
		
		\item Vertices 5-8 correspond to vertices 1-4 \R{but omit} the first constraint in Eqs. \eqref{v1}, \eqref{asf}, \eqref{v3}, and \eqref{sdf}, respectively. In these vertices, there are no specific constraints for the CL. \R{The focus is primarily} on the performance of the closed-loop latency and FL's jitter, which \R{applies to} systems without the control for RA, \R{such as} the authorization of avatar usage rights mentioned in Section I-A.
	\end{itemize}
	
	\R{It is important to obtain} the distributions of $T_{\rm PF1}^{\rm q}$ and $T_{\rm PF2}^{\rm q}$ for modeling vertices 1, 3, 5, and 7. \R{However, to the best of the authors' knowledge, no conclusive results exist for these distributions.}\footnote{\R{Once these distributions are known, vertices 1, 3, 5, and 7 can be analytically modeled using the proposed method in this paper.}} Thus, we focus on the modeling and analysis of vertices 2, 4, 6, and 8 in the following parts, which \R{correspond to} cases where the transmission of PF data has CSIRT and FSD. \R{Given the distribution of the inter-arrival time, the proposed framework and analyzing method can also be extended to other goal-oriented metrics, such as the age of information \cite{Yates2021}, age of loop \cite{aol}, \cite{Cao2023}, and age of incorrect information \cite{Maatouk2020}.} Moreover, large deviation theory can be applied to model the queueing processes at the BS and RA for ET data\cite{li2023arxiv, li2021gc}, \R{capturing the asymptotic distribution of queueing latency. These extensions are left for future work.}

	\section{Stochastic Latency Components and a Unified Optimization Problem in CGC Systems}

	In this section, we will analyze the distribution of communication, computation, compression, and decompression latency in CGC systems. A unified optimization problem is formulated by \R{integrating various constraints and stochastic latency components.}

	\subsection{Distributions of Latency Components in CGC Systems}
	
	We start with the introduction of the communication latency.
	The latency for transmitting a packet, \R{denoted as $t_i$ where $i \in \{{\rm HL},{\rm LL},{\rm PF1}, {\rm PF2},{\rm CD},{\rm VI}\}$, satisfies $t_{i}=\frac{n_i}{R_i}$.} Thus, the total time to successfully transmit a packet follows a geometric distribution. As defined in Section II-A, $T_i^{\rm cm}$ denotes the transmission latency for $N_i$ packets, which follows a negative binomial distribution $\text{NB}(N_i,1-\epsilon_i)$. The PDF of $T^{\rm cm}_i$ is given by
	\begin{equation}
		\Pr\{T^{\rm cm}_i=(N_i+k)t_{i}\}={{k+N_i-1}\choose{N_i-1}}(1-\epsilon_i)^{N_i}\epsilon_i^{k},
	\end{equation}
	where $k\in\{0,1,\cdots\}$.
	
	PF data is collected every $T_{\rm s}$ seconds. Thus, a packet containing the PF data arrives at the buffer every $T_{\rm s}$ seconds, \R{which leads to queueing delays}.	For vertices 2, 4, 6, and 8, we consider the case where CSIRT and FSD are available for the transmission of PF data. With CSIRT and FSD, the deadline constraint can be \R{satisfied using} finite average power \cite{twc1}. Specifically,  for $j\in\{{\rm PF1},{\rm PF2}\}$, the deadline-constrained capacity is approximated by \cite{icc2024,iot1}
	\begin{equation}\label{dcc}
		R_{j}=B_{j}\log_2\left(1+\frac{c_1d_{j}^{-\ell_{j}}P_{j}}{N_0B_{j}\left(c_2+(\tau_{j})^{-c_3}\right)}\right), 
	\end{equation}
	 where $c_i>0$, $i\in\{1,2,3\}$ are constants. 
	\R{For BS-to-UE transmission, the arrival of PF data is not periodic, but it can be regularized by reintegration, i.e., allowing one packet into the transmitting buffer even if multiple packets arrive in the same time slot.} Therefore, the conclusion is valid.
	According to our previous conclusions in \cite{icc2024} and \cite{iot1}, by transmitting the PF data through two antennas with maximum-ratio combining, the closed-form approximation for the deadline-constrained capacity is 
	\begin{equation}\label{aphd}
		R_{j}=B_{j}\log_2\left(1+\frac{2.8771d_{j}^{-\ell_{j}}P_{j}}{N_0B_{j}\left(1.8771+\left(\tau_{j}\right)^{-3.411}\right)}\right).
	\end{equation}
	By this means, to ensure that all PF data is transmitted within $\tau_{j}$, the average power $P_{j}$ can be determined from Eq. \eqref{aphd} by setting \R{$R_{j}= \frac{n_{\rm PF}}{T_{\rm s}}$}.

	\R{Next,} we introduce the formulation of the computation time.  As shown in Fig. \ref{bsysmod}, there are two feature extraction procedures in the CGC system, which are \R{computed} at the MEC server. For $l\in\{{\rm LL},{\rm VI}\}$, the computation time is given by
	\begin{equation}
		T^{\rm cp}_{l}=\frac{\tilde{n}_lX^{\rm cp}_l}{\chi_{\rm MEC}},
	\end{equation}
	where $\chi_{\rm MEC}$ is the frequency of the processing unit at the MEC server, and $X^{\rm cp}_l$ is the number of processing-unit cycles required for processing 1-bit data. Here, $\tilde{n}_{\rm LL}=N_{\rm HL}n_{\rm HL}$ and $\tilde{n}_{\rm VI}=n_{\rm ET}$. According to prior research \cite{lorch2001, suman2023,han2019}, the number of cycles
	allocated to compute 1 bit is stochastic in nature. For the CPU case, $X^{\rm cp}_l$ follows a gamma distribution ${\rm Gamma}(\alpha_{\rm MEC},\beta_l)$ \cite{lorch2001, suman2023,han2019}, where $\alpha_{\rm MEC}$ and $\beta_l$ are the shape and rate parameters of the gamma distribution, respectively. The shape parameter remains
	the same for all tasks performed by the same processor, while the scale parameter $\beta_l$ will be different for
	different tasks \cite{suman2023}. Consequently, the computation time follows a gamma distribution ${\rm Gamma}\left(\alpha_{\rm MEC},\frac{\chi_{\rm MEC}\beta_l}{\tilde{n}_l}\right)$. The PDF of $T^{\rm cp}_l$ is given by
	\begin{equation}
		\!\! f_{l}^{\rm cp}(x)\!=\!\frac{\left(\chi_{\rm MEC}\beta_l\right)^{\alpha_{\rm MEC}}}{\tilde{n}_l^{\alpha_{\rm MEC}}\Gamma(\alpha_{\rm MEC})}x^{\alpha_{\rm MEC}-1}\exp\left(\!-\frac{\chi_{\rm MEC}\beta_l}{\tilde{n}_l}x\!\right), 
	\end{equation} 
	where $x>0$. The expectation of $T^{\rm cp}_l$ is $\mu^{\rm cp}_l=\frac{\tilde{n}_l\alpha_{\rm MEC}}{\chi_{\rm MEC}\beta_l}$, while its variance is $\xi^{\rm cp}_l=\frac{\tilde{n}_l^2\alpha_{\rm MEC}}{\left(\chi_{\rm MEC}\beta_l\right)^2}$. \footnote{Note that the analysis in this paper is also valid for GPU scenarios if the distribution of $X^{\rm cp}_l$ for GPU is given.} 
	
	For the compression time, we \R{define $\kappa\in [1,\kappa_{\rm max}]$ as} the compression ratio, which is the ratio of the size of raw data to that of compressed data. \R{Here, $\kappa_{\rm max}$ denotes} the maximum compression ratio. Let $X_{\rm c}$ denote the number of CPU cycles \R{required for} compressing 1-bit data, which follows ${\rm Gamma}\left(\alpha_{\rm RA},\beta_{\rm c}\right)$, where $\alpha_{\rm RA}$ and $\beta_{\rm c}$ denote the shape and rate parameters of compression latency, respectively. \R{Let $\zeta_{\rm c}(\kappa)=\mathbb{E}\{X_{\rm c}\}$, which is an increasing function of $\kappa$. There are two primary conclusions in the literature for evaluating the expected computation load given a specific compression ratio. However, it remains unclear which conclusion is more applicable and what the exact range of application for each is. Therefore, we consider both conclusions to enhance the comprehensiveness of our analysis. We use $X_{\rm c}^{(1)}$ and $X_{\rm c}^{(2)}$ to distinguish between these two different conclusions.} The first conclusion shows that
	\begin{equation}\label{cpm0}
		\R{\mathbb{E}\left\{X^{(1)}_{\rm c}\right\}=\frac{\alpha_{\rm RA}}{\beta_{\rm c}}=\exp\left(\psi \kappa\right)-\exp\left(\psi\right),}
	\end{equation}
	where $\psi>0$ is a constant. This model can be used for several lossless compression techniques such as Zlib, Zstandard, and XZ compression \cite{suman2023,han2019,li2019}. Another conclusion shows that 
	\begin{equation}\label{cpm1}
		\R{\mathbb{E}\left\{X^{(2)}_{\rm c}\right\}=\frac{\alpha_{\rm RA}}{\beta_{\rm c}}=\omega_1\left(\omega_2 \kappa^{\omega_3}+\omega_4\right),}
	\end{equation}
	where $\omega_k$, $k=1,\ldots,4$ are positive constants for lossless compression algorithms GZIP and BZ2 as shown in \cite{nguyen2020}.\footnote{Note that this model can also be used for evaluating $\mathbb{E}\{X_{\rm C}\}$ of the lossy compression \cite{nguyen2020}, e.g., JPEG.} \R{According to \cite{nguyen2020}, Eq. \eqref{cpm1} is also applicable to Zlib and XZ compression. Most of our analyses do not depend on the form of $\zeta_{\rm c}(\kappa)$. For some discussions in Sections IV-C, IV-D, and V, we will specify which form we consider for $\zeta_c(\kappa)$. }
	
		Let $X_{\rm d}$ denote the number of CPU cycles for decompressing 1-bit data. $X_{\rm d}$ follows ${\rm Gamma}\left(\alpha_{\rm MEC}, \beta_{\rm d}\right)$, where $\beta_{\rm d}$ is the rate parameter of decompression latency. \R{Let $\zeta_{\rm d}(\kappa)=\mathbb{E}\left\{X_{\rm d}\right\}$.} Accordingly, there are two primary conclusions in the literature for evaluating $\mathbb{E}\left\{X_{\rm d}\right\}$. \R{We use $X_{\rm d}^{(1)}$ and $X_{\rm d}^{(2)}$ to differentiate between the two different conclusions in existing works, corresponding to $X_{\rm c}^{(1)}$ and $X_{\rm c}^{(2)}$, respectively.} The first conclusion shows that 
	\begin{equation}\label{decom1}
		\R{\mathbb{E}\left\{X_{\rm d}^{(1)}\right\}=\frac{\alpha_{\rm MEC}}{\beta_{\rm d}}=\omega_0\left(\exp\left(\psi \kappa\right)-\exp\left(\psi\right)\right),}
	\end{equation}
	where $\omega_0\in(0,1)$ is a constant. Another conclusion shows 
	\begin{equation}\label{decom2}
		\R{\mathbb{E}\left\{X_{\rm d}^{(2)}\right\}=\frac{\alpha_{\rm MEC}}{\beta_{\rm d}}=\omega_5\left(\omega_6\kappa^{\omega_7}+\omega_8\right),}
	\end{equation}
	where $\omega_5$, $\omega_6$, and $\omega_8$ are positive constants. The sign of $\omega_7$ is determined by the specific compression algorithms \cite{nguyen2020}.  
	
	With the above models and definitions, $T_{\rm ET}^{\rm c}=\frac{n_{\rm ET}X_{\rm c}}{\chi_{\rm RA}}$, and $T_{\rm ET}^{\rm d}=\frac{n_{\rm ET}X_{\rm d}}{\kappa\chi_{\rm MEC}}$, where $\chi_{\rm RA}$ is the frequency of the processing unit at RA. Thus, we obtain $T_{\rm ET}^{\rm c}\sim {\rm Gamma}\left(\alpha_{\rm RA},\frac{\chi_{\rm RA}\alpha_{\rm RA}}{n_{\rm ET}\zeta_{\rm c}(\kappa))}\right)$, with expectation $\mu_{\rm ET}^{\rm c}=\frac{n_{\rm ET}\zeta_{\rm c}(\kappa)}{\chi_{\rm RA}}$ and variance $\xi^{\rm c}_{\rm ET}=\frac{\left(n_{\rm ET}\zeta_{\rm c}(\kappa)\right)^2}{\chi^2_{\rm RA}\alpha_{\rm RA}}$.
	$T_{\rm ET}^{\rm d}\sim{\rm Gamma} \left(\alpha_{\rm MEC},\frac{\chi_{\rm MEC}\alpha_{\rm MEC}\kappa}{n_{\rm ET}\zeta_{\rm d}(\kappa)}\right)$, with expectation $\mu_{\rm ET}^{\rm d}=\frac{n_{\rm ET}\zeta_{\rm d}(\kappa)}{\chi_{\rm MEC}\kappa}$ and variance $\xi_{\rm ET}^{\rm d}=\frac{\left(n_{\rm ET}\zeta_{\rm d}(\kappa)\right)^2}{\left(\chi_{\rm MEC}\kappa\right)^2\alpha_{\rm MEC}}$.

	\subsection{Formulation of a Unified Optimization Problem }
	
	In this subsection, we will formulate a unified optimization problem for this CGC system, in which average-based and tail-based optimization objectives and constraints are considered.
	
	\R{The optimization objective can be framed from various aspects.} If \R{our goal is} to reduce the closed-loop latency to enhance reliability and provide a better experience for users, the objective function is related to the closed-loop latency $T$. If we aim to save resources, e.g., power and bandwidth, the objective function is related to the average power and bandwidth.

	For the first kind of objective functions, we also consider two cases, i.e., the average-based and tail-based objectives. The average-based objective is associated to $\mathbb{E}\{T\}$. For the tail-based objective, let $T_{\rm th}$ denote the threshold for the closed-loop latency, which is determined by the specific applications and user's requirements. Thus, the tail-based objective is related to $\Pr\{T<T_{\rm th}\}$. We \R{define $d(T)$ as the objective function associated with $T$}. Besides, we \R{define $g(P_{\rm avg},B)$ as the objective function related to resources}. We also assume that $d(T)$ and $g(P_{\rm avg},B)$ are normalized.
	
	A unified optimization problem for this CGC system is then formulated by combining different kinds of objectives and constraints.
	\begin{subequations}\label{optpro}
		\begin{align}
			\min \quad & \varepsilon d(T)+(1-\varepsilon)g\left(P_{\rm avg},B\right) \\
			\mbox{s.t.}\quad
			& \text{Constraints of a vertex in Fig. \ref{constraints},} \label{add5}\\
			&P_{\rm PF1}+P_{\rm CD}\leq P_{\rm RA}^{\rm th},\label{conc}\\
			&P_{\rm PF2}+P_{\rm VI}\leq P_{\rm BS}^{\rm th}, \label{cond}\\
			&B_{\rm PF}+B_{\rm ET}= B,
		\end{align}
	\end{subequations}
	where $\varepsilon\in[0,1]$, $P_{\rm RA}^{\rm th}$ and $P_{\rm BS}^{\rm th}$ \R{denote the average power constraints} for the RA and BS, respectively. \R{We set $B_{\rm PF1}=B_{\rm PF2}=B_{\rm PF}$, and $B$ (in Hz) represents the maximum available bandwidth.} \R{Eq. \eqref{add5} allows us to select a vertex based on scenario requirements, with each vertex corresponding to the specific constraints outlined in Section II-D.} For $\varepsilon=0$, we are interested in designing a resource-efficient scheduling. For $\varepsilon=1$, \R{the focus is on} optimizing the performance of closed-loop latency.
	The optimization variables in this problem may include $\left\{\kappa,B_{\rm PF},B_{\rm ET},P_{\rm PF1},P_{\rm PF2},\tau_{\rm PF1}, \tau_{\rm PF2}, P_{\rm CD},P_{\rm VI}\right\}$.

	It is not trivial work to solve this optimization problem in CGC systems. As \R{demonstrated} in Section III-A, the expectation and variance of each latency component in this system are known, which can be used to characterize some constraints related to the expectation and variance of $T_{\rm ET}$. However, \R{regardless of the vertex constraints chosen}, we have to obtain the distribution of closed-loop and FL’s latency. For vertices 2 and 4, \R{the CDF of CL latency is also required}. Since the latency of CL, FL, and the closed loop are sums of gamma and negative binomial random variables, \R{obtaining analytical distributions for these latency is challenging.} \R{Thus, it is crucial to derive the analytical expression for the distribution of CL, FL, and closed-loop latency in general scenarios for enabling efficient performance evaluation or optimization.}	

	\begin{figure}[t]
		\centerline{\includegraphics[width=8.5cm]{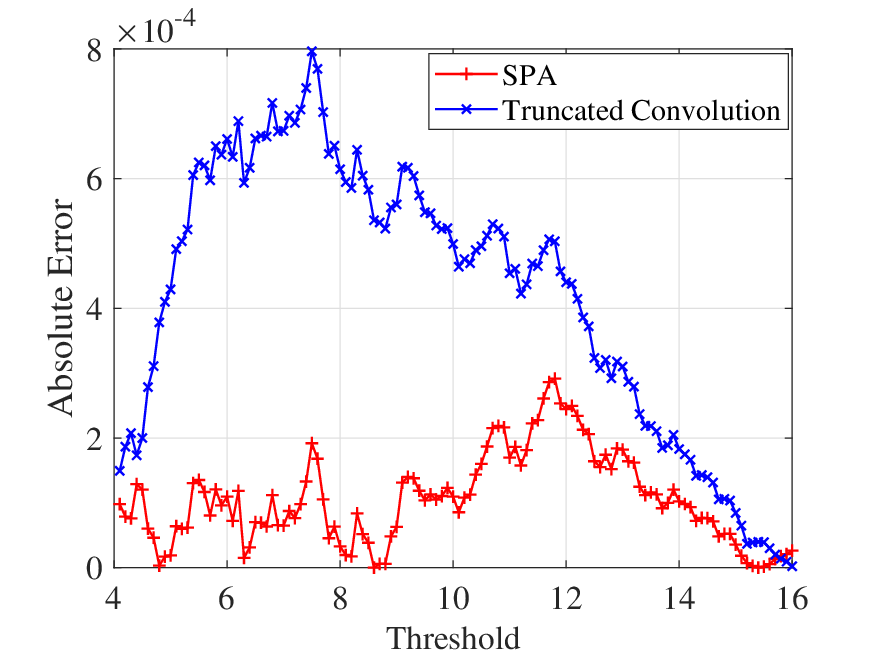}}
		\caption{\R{Comparison between the SPA and the truncated convolution used in  \cite{suman2023}.}}
		\label{compare}
	\end{figure}

	\section{Analytical Expression of the Latency Distribution in CGC Systems}

	In this section, we will present a general method for approximating the distribution of CL, FL, and closed-loop latency. Based on this general method, we also provide methods for efficiently evaluating the terms in various constraints. Since there are similarities in the latency distributions and constraints of CL, FL, and closed-loop latency, we will take FL for example to show how to derive the analytical expression for the terms in the proposed constraints for FL. \R{We will then present} the results of CL and closed-loop latency without detailed proofs. For simplicity, we define $T_1=T_{\rm CL}+T_{\rm ET}$. Without loss of generality, we let $t_{\rm CD}=t_{\rm VI}=t_{\rm u}$.

	\begin{figure*}[!b]
		\normalsize 
		\hrulefill
		\newcounter{TempEqCnt}
		\setcounter{TempEqCnt}{\value{equation}} 
		\setcounter{equation}{30} 
		\begin{equation}\label{cgftmd}
			\begin{aligned}
				L(s)=-\alpha_{\rm RA}\ln\left(1-\frac{n_{\rm ET}\zeta_{\rm c}(\kappa)s}{\chi_{\rm RA}\alpha_{\rm RA}}\right)-\alpha_{\rm MEC}\ln\left(1-\frac{n_{\rm ET}\zeta_{\rm d}(\kappa)s}{\chi_{\rm MEC}\alpha_{\rm MEC}\kappa}\right)-\alpha_{\rm MEC}\ln\left(1-\frac{n_{\rm ET}s}{\chi_{\rm MEC}\beta_{\rm VI}}\right).
			\end{aligned}
		\end{equation}
	\end{figure*}

	\subsection{A General Method for Analytical Approximation for the CDF of Latency}
	
From Section III, we find that there are both discrete and continuous random variables among the components of CL, FL, and closed-loop latency. We have to treat them separately \R{due to their different properties}. We will first present a general method to approximate the CDF of $T_{\rm ET}$, which can be applied to the CDF \R{approximations} of $T_{\rm CL}$, $T_{\rm FL}$, and $T$.

	\R{In Lemma 1, we derive the analytical expression for the CDF of $T^{\rm c}_{\rm ET}+T^{\rm d}_{\rm ET}+T^{\rm cp}_{\rm VI}$ with the help of SPA.} For simplicity, let $F_{1}(x)$ and $f_1(x)$ denote the CDF and PDF, respectively, of $T^{\rm c}_{\rm ET}+T^{\rm d}_{\rm ET}+T^{\rm cp}_{\rm VI}$, \R{which is a sum of continuous random variables.} Besides, we let 
		\setcounter{equation}{25}
	\begin{equation}\label{const}
		\iota_1=\left(\frac{n_{\rm ET}^2\zeta_{\rm c}^2(\kappa)}{\chi_{\rm RA}^2 \alpha_{\rm RA}}+\frac{n_{\rm ET}^2\zeta^2_{\rm d}(\kappa)}{\chi_{\rm MEC}^2 \kappa^2 \alpha_{\rm MEC}}+\frac{n_{\rm ET}^2 \alpha_{\rm MEC}}{\chi^2_{\rm MEC}\beta_{\rm VI}^2}\right)^{\frac{3}{2}},
	\end{equation}
	\begin{equation}\label{const1}
		\iota_2=\frac{2n_{\rm ET}^3\zeta_{\rm c}^3(\kappa)}{\chi_{\rm RA}^3\alpha_{\rm RA}^2}+\frac{2n_{\rm ET}^3\zeta_{\rm d}^3(\kappa)}{\kappa^3\chi_{\rm MEC}^3\alpha_{\rm MEC}^2}+\frac{2\alpha_{\rm MEC}n_{\rm ET}^3}{\chi_{\rm MEC}^3\beta_{\rm VI}^3}, \, \text{and}
	\end{equation}
	\begin{equation}\label{expectmdc}
		\begin{aligned}
			\theta=\frac{n_{\rm ET}\alpha_{\rm MEC}}{\chi_{\rm MEC}\beta_{\rm VI}}+\frac{n_{\rm ET}\zeta_{\rm c}(\kappa)}{\chi_{\rm RA}}&+\frac{n_{\rm ET}\zeta_{\rm d}(\kappa)}{\chi_{\rm MEC}\kappa}.
		\end{aligned}
	\end{equation}

	\textbf{Lemma 1}. With $\theta$ defined in Eq. \eqref{expectmdc}, 
	\begin{itemize}
		\item the CDF of $T^{\rm c}_{\rm ET}+T^{\rm d}_{\rm ET}+T^{\rm cp}_{\rm VI}$ is approximated by
		\begin{equation}\label{lemma11}
			\R{\begin{aligned}
					F_{1}(x)\approx\begin{cases}
						\Phi(v_x)+\phi(v_x)\left ( \frac{1}{v_x}-\frac{1}{u_x} \right),  \quad &x\neq \theta,\\
						\frac{1}{2}+\frac{\iota_2}{6\sqrt{2\pi }\iota_1}, \quad & x=\theta,
					\end{cases}
			\end{aligned}}
		\end{equation}
		where
		\begin{equation}\label{uv}
			\begin{aligned}
				v_x&={\rm sign}\left(\Xi_1\left(x\right)\right)\sqrt{2x\Xi_1\left(x\right)-2L(\Xi_1(x))},\,  \text{and} \\
				u_x&=\Xi_1(x)\sqrt{L''(\Xi_1(x))}.
			\end{aligned}
		\end{equation}
		$\Xi_1(\cdot)$ is the inverse function of $L'(\cdot)$. $\iota_1$ and $\iota_2$ are given in Eqs. \eqref{const} and \eqref{const1}, respectively. $L(\cdot)$ is given in Eq. \eqref{cgftmd}, where $s<\min\left\{ \frac{\chi_{\rm RA}\alpha_{\rm RA}}{n_{\rm ET}\zeta_{\rm c}(\kappa)}, \frac{\chi_{\rm MEC}\alpha_{\rm MEC}\kappa}{n_{\rm ET}\zeta_{\rm d}(\kappa)},   \frac{\chi_{\rm MEC}\beta_{\rm VI}}{n_{\rm ET}}   \right\}$. 
		
		\item the PDF of $T^{\rm c}_{\rm ET}+T^{\rm d}_{\rm ET}+T^{\rm cp}_{\rm VI}$ is approximated by
		\setcounter{equation}{31}
		\begin{equation}\label{lemma13}
			\R{	f_{1}(x)\approx\frac{\exp\left(L(\Xi_1(x))-x\Xi_1(x)\right)}{\sqrt{2\pi L''(\Xi_1(x))}}, \, x>0.}
		\end{equation}
	\end{itemize}

	\begin{IEEEproof}
		See Appendix A.
	\end{IEEEproof}

	In Lemma 1, we derive the CDF and PDF of $T^{\rm c}_{\rm ET}+T^{\rm d}_{\rm ET}+T^{\rm cp}_{\rm VI}$, which is the sum of continuous components of $T_{\rm ET}$. For $T_1$, the sum of continuous components is $T_{\rm LL}^{\rm cp}+T^{\rm c}_{\rm ET}+T^{\rm d}_{\rm ET}+T^{\rm cp}_{\rm VI}$. Thus, we define $\tilde{L}(s)$ as 
	\begin{equation}\label{tt1}
		\tilde{L}(s)=L(s)-\alpha_{\rm MEC}\ln\left(1-\frac{N_{\rm HL}n_{\rm HL}s}{\chi_{\rm MEC}\beta_{\rm LL}}\right),
	\end{equation}
	where $s<\min\left\{  \frac{\chi_{\rm MEC}\beta_{\rm LL}}{N_{\rm HL}n_{\rm HL}} , \frac{\chi_{\rm RA}\alpha_{\rm RA}}{n_{\rm ET}\zeta_{\rm c}(\kappa)}, \frac{\chi_{\rm MEC}\alpha_{\rm MEC}\kappa}{n_{\rm ET}\zeta_{\rm d}(\kappa)},   \frac{\chi_{\rm MEC}\beta_{\rm VI}}{n_{\rm ET}}   \right\}$. Besides, we let $\tilde{\iota}_1=\left(\iota_1^{\frac{2}{3}}+\frac{N_{\rm HL}^2n_{\rm HL}^2\alpha_{\rm MEC}}{\chi^2_{\rm MEC}\beta_{\rm LL}^2}\right)^{\frac{3}{2}}$, $\tilde{\iota_2}=\iota_2+\frac{2\alpha_{\rm MEC}N^3_{\rm HL}n_{\rm HL}^3}{\chi^3_{\rm MEC}\beta^3_{\rm LL}}$, and $\tilde{\theta}=\theta+\frac{N_{\rm HL}n_{\rm HL}\alpha_{\rm MEC}}{\chi_{\rm MEC}\beta_{\rm LL}}$. By replacing $\tilde{L}(s)$, $\tilde{\iota}_1$, $\tilde{\iota}_2$, and $\tilde{\theta}$ into Eqs. \eqref{lemma11} and \eqref{lemma13}, we obtain the CDF and PDF of $T_{\rm LL}^{\rm cp}+T^{\rm c}_{\rm ET}+T^{\rm d}_{\rm ET}+T^{\rm cp}_{\rm VI}$. We let $\tilde{F}_1(x)$ denote the CDF of $T_{\rm LL}^{\rm cp}+T^{\rm c}_{\rm ET}+T^{\rm d}_{\rm ET}+T^{\rm cp}_{\rm VI}$.

	\R{To show the advantages of SPA, we provide a comparison between SPA and the truncated convolution used in \cite{suman2023}. The comparison result is shown in Fig. \ref{compare}. We consider the sum of $M=4$ independent gamma random variables, each following ${\rm Gamma}(2,1)$. The granularity of the truncated convolution is set to $\delta_1=10^{-3}$, and the truncated threshold is defined as $\Lambda_1=20$. The solution to the SPA equation is obtained through binary search with termination threshold $\delta_2= 10^{-3}$ and an interval $[\Lambda_2,\Lambda_3]$, where $\Lambda_2=-10^{5}$ and $\Lambda_3=1$. For SPA, the granularity is set to $\delta_3=10^{-3}$, and the truncated threshold is $\Lambda_4=16$. From Fig. \ref{compare}, we observe that SPA achieves a smaller error than the truncated convolution in most cases. The complexity of the truncated convolution is $O\left(\left\lceil\frac{\Lambda_1}{\delta_1}\right\rceil^M\right)$ for the general cases where random variables are independent but not necessarily identically distributed. In contrast, the complexity of SPA is $O\left(\left\lceil \frac{\Lambda_4}{\delta_3}\right \rceil \log_2 \left\lceil \frac{\Lambda_3-\Lambda_2}{\delta_2}\right\rceil \right)$, which is not related to $M$. Thus, SPA offers improved approximation accuracy while reducing the complexity for larger $M$.}

	\R{\emph{Remark 1: The existing work has discussed the asymptotic error of SPA for the CDF of the sum of i.i.d. random variables. For i.i.d. random variables $Y_1, Y_2, \cdots, Y_n$, let $Y=\sum_{i=1}^n Y_i$, and let $\tilde{F}_Y(y)$ denote the SPA of its CDF, while $F_Y(y)$ denotes the true CDF of $Y$. The error of the SPA for CDF mainly results from the Laplace approximation \cite[Eq. (2.5)]{LR} and Temme approximation \cite[Eq. (2.34)]{LR}. According to \cite[Eq. (2.46)]{LR}, the relationship between $\tilde{F}_Y(y)$ and $F_Y(y)$ is given by }}
	\begin{equation}\label{add1}
		\R{F_Y(y)=\tilde{F}_Y(y)+O\left(n^{-1}\right).}
	\end{equation}

\R{\emph{This error analysis can be viewed as a special case of the considered problem in this paper. For example, when $T_{\rm ET}^{\rm c}$, $T_{\rm ET}^{d}$, $T_{\rm VI}^{\rm cp}$ are i.i.d. in Lemma 1, Eq. \eqref{add1} holds. However,  $T_{\rm ET}^{\rm c}$, $T_{\rm ET}^{d}$, $T_{\rm VI}^{\rm cp}$ may not follow identical gamma distributions. For this case, we can also utilize the result of Eq. (2.46) in \cite{LR} to provide an intuitive analysis, though not rigorous. Assume that $Y_i$ follows ${\rm Gamma}(\alpha_i,\beta_i)$ and $Y_i$ is independent of $Y_j$, $i\neq j$. For convenience of expression, we let $\upsilon_1=\max_{i=1,\cdots,n}\{\alpha_i\}$ and $\upsilon_2=\min_{i=1,\cdots,n}\{\beta_i\}$. Then, the CGF of $\sum_{i=1}^{n} Y_i$ can be expressed as}}
\begin{equation}\label{add2}
	\R{\begin{aligned}
			K_Y(s)&=-\sum_{i=1}^n \alpha_i\ln\left(1-\frac{s}{\beta_i}\right) \\
			&=n\underbrace{\left[-\upsilon_1\ln\left(1-\frac{s}{\upsilon_2}\right)\right]\left[\sum_{i=1}^{n}\frac{\alpha_i}{n\nu_1}\frac{\ln\left(1-\frac{s}{\beta_i}\right)}{\ln\left(1-\frac{s}{\upsilon_2}\right)}\right]}_{K_{\hat{Y}}(s)},\\
			&=n\left[-\upsilon_1\ln\left(1-\frac{s}{\upsilon_2}\right)\right]\left(1+O(n^{-1})\right), \quad s<\upsilon_2
	\end{aligned}}
\end{equation}
\R{\emph{Eq. \eqref{add2} holds because $\frac{\alpha_i}{\upsilon_1}<1$ and $\frac{\ln\left(1-\frac{s}{\beta_i}\right)}{\ln\left(1-\frac{s}{\upsilon_2}\right)}$ is bounded. The boundness of  $\frac{\ln\left(1-\frac{s}{\beta_i}\right)}{\ln\left(1-\frac{s}{\upsilon_2}\right)}$ can be shown by $\lim\limits_{s\to\upsilon_2} \frac{\ln\left(1-\frac{s}{\beta_i}\right)}{\ln\left(1-\frac{s}{\upsilon_2}\right)}=0$ with $\beta_i\neq \upsilon_2$, $\lim\limits_{s\to\upsilon_2} \frac{\ln\left(1-\frac{s}{\beta_i}\right)}{\ln\left(1-\frac{s}{\upsilon_2}\right)}=1$ with $\beta_i= \upsilon_2$, and}}
\begin{equation}
	\R{\lim_{s\to-\infty} \frac{\ln\left(1-\frac{s}{\beta_i}\right)}{\ln\left(1-\frac{s}{\upsilon_2}\right)}=\lim_{s\to-\infty}\frac{\ln\left(-\frac{s}{\beta_i}\right)}{\ln\left(-\frac{s}{\upsilon_2}\right)}=1.}
\end{equation}
\R{\emph{Eq. \eqref{add2} indicates that we can treat $\sum_{i=1}^n Y_i$ as a summation of $n$ i.i.d. random variables with CGF $K_{\hat{Y}}(s)$ with error $O(n^{-1})$. The essence of this analysis is that the CGF of $Y_i$ has a unified expression with different values of parameters.  Combining Eqs. \eqref{add2} and \eqref{add1}, we can approximately evaluate the error of SPA in Lemma 1. However, it is an important work to provide rigorous error analysis for SPA with non-i.i.d. random variables, which remains open and is beyond the scope of this paper. From Eq. \eqref{add1}, we find that SPA becomes more accurate with a larger $n$. However, as noted in \cite{LR} and demonstrated by the numerical results in Section V, SPA achieves satisfying accuracy with $n=3$. To enhance accuracy, we can adopt the second-order SPA, which can be found in \cite[Eq. (2.48)]{LR} with an error term $O\left(n^{-2}\right)$. For the SPA of the PDF, the analysis is similar. We will elaborate on the error analysis for the SPA of the PDF in Remark 2, which is analogous to the error analysis for the PMF approximation.}}

	\begin{figure*}[b]
		\normalsize \hrulefill
				\setcounter{equation}{37} 
		\begin{equation}\label{cgfdis1}
			J(s)\!=\!N_{\rm CD}\left(s+
			\ln(1-\epsilon_{\rm CD})-\ln(1-\epsilon_{\rm CD}e^s)\right)\!+\!N_{\rm VI}\left(s+\ln(1-\epsilon_{\rm VI})-\ln\left(1-\epsilon_{\rm VI}e^s\right)\right), \, s\!<\!\min \left\{-\ln\epsilon_{\rm CD},-\ln\epsilon_{\rm VI} \right\}.
		\end{equation}	
		\setcounter{equation}{39} 
		\begin{equation}\label{add4}
			\R{\begin{aligned}
					K_Z(s)=&n\underbrace{ \upsilon_3(s+\ln(1-\upsilon_5)-\ln(1-\upsilon_4e^s))\left(\sum_{i=1}^n\frac{K_{Z_i}(s)}{n\upsilon_3(s+\ln(1-\upsilon_5)-\ln(1-\upsilon_4e^s))}\right)}_{K_{\hat{Z}}(s)}\\
					&= n \upsilon_3(s+\ln(1-\upsilon_5)-\ln(1-\upsilon_4e^s))\left(1+O(n^{-1})\right), \quad \quad \quad \quad \quad s<-\ln \upsilon_4.
			\end{aligned}}
		\end{equation}
	\end{figure*}
	
	In Lemma 2, we obtain the closed-form expression for the probability mass function (PMF) of $T^{\rm cm}_{\rm CD}+T^{\rm cm}_{\rm VI}$ with the help of saddlepoint mass functions. \R{$T^{\rm cm}_{\rm CD}+T^{\rm cm}_{\rm VI}$ is a sum of discrete components of $T_{\rm ET}$.} For simplicity, let $p_k$ denote $\Pr\left\{T^{\rm cm}_{\rm CD}+T^{\rm cm}_{\rm VI}=kt_{\rm u}\right\}$, where $k\in\{N_{\rm CD}+N_{\rm VI},N_{\rm CD}+N_{\rm VI}+1,\cdots\}$.

	\textbf{Lemma 2} The PMF of $T^{\rm cm}_{\rm CD}+T^{\rm cm}_{\rm VI}$ is approximated by
	\setcounter{equation}{36} 
	\begin{equation}\label{lemma21}
		\begin{aligned}
			\R{p_k\approx\begin{cases}
					(1-\epsilon_{\rm CD})^{N_{\rm CD}}(1-\epsilon_{\rm VI})^{N_{\rm VI}}, \quad k=N_{\rm CD}+N_{\rm VI},\\
					\frac{\exp\left(J(\Xi_2(k))-k\Xi_2(k)\right)}{\sqrt{ 2\pi\left( \frac{\epsilon_{\rm CD}N_{\rm CD}e^{\Xi_2(k)}}{\left(1-\epsilon_{\rm CD}e^{\Xi_2(k)}\right)^2}+\frac{\epsilon_{\rm VI}N_{\rm VI}e^{\Xi_2(k)}}{\left(1-\epsilon_{\rm VI}e^{\Xi_2(k)}\right)^2}\right)}},  \\ \quad \quad \quad \quad \quad \quad \quad\quad \quad \quad \quad \quad \,\, k>N_{\rm CD}+N_{\rm VI},
			\end{cases}}
		\end{aligned}
	\end{equation}
	where $\Xi_2(\cdot)$ is the inverse function of $J'(\cdot)$. $J(\cdot)$ is given in Eq. \eqref{cgfdis1}.

	\begin{IEEEproof}
		See Appendix B.
	\end{IEEEproof}
	
	\R{\emph{Remark 2: The existing work has conducted the error analysis for saddlepoint mass function with i.i.d. random variables $Z_i$, $i=1,\cdots,n$. Let $\{P_k\}$ denote the true PMF of $Z=\sum_{i=1}^n Z_i$, while $\{\tilde{P}_k\}$ denotes the saddlepoint mass function. Assuming that $Z_i$ is i.i.d., we have \cite[Eq. (2.24)]{LR}}}
	\setcounter{equation}{38} 
	\begin{equation}\label{add3}
		\R{	P_k=\tilde{P}_k(1+O(n^{-1})).}
	\end{equation}
	\R{\emph{Eq. \eqref{add3} can be used to analyze the approximation error in Lemma 2 when $T_{\rm CD}^{\rm cm}$ and $T_{\rm VI}^{\rm cm}$ are i.i.d.. When $T_{\rm CD}^{\rm cm}$ and $T_{\rm VI}^{\rm cm}$ are independent but not identically distributed, the analysis is similar to Remark 1. Assume that $Z_i\sim {\rm NB}(N_i,1-\epsilon_i)$ and $Z_i$ is independent of $Z_j$ with $i\neq j$. The CGF of $Z_i$ is denoted by $K_{Z_i}(s)$. Let $\upsilon_3=\max\{N_i\}$, $\upsilon_4=\max\{\epsilon_i\}$, and $\upsilon_5=\min\{\epsilon_i\}$. Then, the CGF of $Z$ can be written as Eq. \eqref{add4}. Eq. \eqref{add4} indicates that we can treat $Z$ as a summation of $n$ i.i.d. random variables with CGF $K_{\hat{Z}}(s)$. Thus, Eq. \eqref{add3} can be used for the error analysis of Lemma 2. To achieve higher approximation accuracy, second-order SPA can be utilized, which is provided in \cite[2.32]{LR}. }}

	In Lemma 2, we derive the PMF of $T^{\rm cm}_{\rm CD}+T^{\rm cm}_{\rm VI}$, which is the sum of discrete components of $T_{\rm ET}$. For $T_1$, the sum of discrete components is $T_{\rm HL}^{\rm cm}+T_{\rm LL}^{\rm cm}+T^{\rm cm}_{\rm CD}+T^{\rm cm}_{\rm VI}$. We let $\tilde{p}_k$ denote $\Pr\{T_{\rm HL}^{\rm cm}+T_{\rm LL}^{\rm cm}+T^{\rm cm}_{\rm CD}+T^{\rm cm}_{\rm VI}=kt_{\rm u}\}$. Moreover, we let $\tilde{J}(s)$ denote the CGF of $T_{\rm HL}^{\rm cm}+T_{\rm LL}^{\rm cm}+T^{\rm cm}_{\rm CD}+T^{\rm cm}_{\rm VI}$. Due to space limits, we omit the expression of $\tilde{J}(s)$, which is similar to Eq. \eqref{cgfdis1}. By substituting $\tilde{J}(s)$ into Eq. \eqref{lemma21}, we obtain $\tilde{p}_k$, where $k>N_{\rm HL}+N_{\rm LL}+N_{\rm CD}+N_{\rm VI}$. For $k= N_{\rm HL}+N_{\rm LL}+N_{\rm CD}+N_{\rm VI}$, we have 
		\setcounter{equation}{40} 
	\begin{equation}
		\tilde{p}_k=(1-\epsilon_{\rm HL})^{N_{\rm HL}}(1-\epsilon_{\rm LL})^{N_{\rm LL}}p_{N_{\rm CD}+N_{\rm VI}}.
	\end{equation}
	
	Moreover, we also let $p^{\rm CL}_k$ denote the PMF of $T_{\rm HL}^{\rm cm}+T_{\rm LL}^{\rm cm}$, which is the sum of discrete component of $T_{\rm CL}$. Due to space limits, we omit the derivation of $p^{\rm CL}_k$, which is also similar to Eq. \eqref{lemma21}.

	Based on Lemmas 1 and 2, \R{CDFs of $T_{\rm CL}$, $T_{\rm ET}$, $T_{\rm FL}$, $T_1$ and $T$ can be derived for general scenarios.} We let $F_{\rm CT}(x)$, $F_{\rm ET}(x)$, $F_{\rm FL}(x)$, $F_{{\rm T}_1}(x)$ and $F_{\rm T}(x)$ denote the CDFs of $T_{\rm CL}$, $T_{\rm ET}$, $T_{\rm FL}$, $T_1$, and $T$, respectively.

	\textbf{Theorem 1}. For vertices 2, 4, 6, and 8, the CDF of $T_{\rm CL}$ is given by
	\begin{equation}
		\R{F_{\rm CL}(x)\!\!\approx\!\!\!\!\sum_{k=N_{\rm HL}+N_{\rm LL}}^{\left\lfloor \frac{x}{t_{\rm u}} \right\rfloor}\!\!\!\!\frac{p_k^{\rm CL}}{\Gamma(\alpha_{\rm MEC})}\gamma\left(\alpha_{\rm MEC},\frac{\chi_{\rm MEC}\beta_{\rm LL}(x-kt_{\rm u})}{N_{\rm HL}n_{\rm HL}}\right),}
	\end{equation}
	where $x>0$.
	
	The CDF of $T_{\rm ET}$ is given by
	\begin{equation}\label{theo11}
		\begin{aligned}
			\R{F_{\rm ET}(x)\approx\sum^{\left \lfloor \frac{x}{t_{\rm u}}\right \rfloor}_{k=N_{\rm CD}+N_{\rm VI}}p_kF_1(x-kt_{\rm u}), \quad x>0.}
		\end{aligned}
	\end{equation}
	
	The CDF of $T_{\rm FL}$ is given by
	\begin{equation}
		\R{F_{\rm FL}(x)\approx\mathbb{I}\left\{x>\tau_{\rm PF}\right\}F_{\rm ET}(x), \quad x>0.}
	\end{equation}
	
	The CDF of $T$ is given by

	\begin{equation}
		\R{F_{\rm T}(x)\approx\mathbb{I}\{x>\tau_{\rm PF}\}F_{\rm CL}(x-\tau_{\rm PF})F_{{\rm T}_1}(x),}
	\end{equation}
	where $F_{{\rm T}_1}(x) $ is given by
	\begin{equation}
		\R{\!\!\! F_{{\rm T}_1}(x)\approx\!\!\!\sum^{\left \lfloor \frac{x}{t_{\rm u}}\right \rfloor}_{k=N_{\rm HL}+N_{\rm LL}+N_{\rm CD}+N_{\rm VI}}\!\! \tilde{p}_k\tilde{F}_1(x-kt_{\rm u}), \quad x>0.}
	\end{equation}
	\begin{IEEEproof}
		The CDF of $T_{\rm CL}$, $T_{\rm ET}$, and $T_{1}$ can be obtained directly according to Lemmas 1 and 2. For $T_{\rm FL}$, we can obtain its CDF based on $F_{\rm ET}(x)$ and the definition of $T_{\rm FL}$. For the closed-loop latency $T$, it can be expressed as $T=T_{\rm CL}+T_{\rm FL}=\max\{T_{\rm CL}+\tau_{\rm PF},T_{1}\}$. Thus, the CDF of $T$ is given by  
		\begin{equation}
			\Pr\{T<x\}=\Pr\{T_{\rm CL}<x-\tau_{\rm PF}\}\Pr\{T_1<x\}.
		\end{equation}
		Then we obtain the CDF of $T$.
	\end{IEEEproof}

	In Theorem 1, we derive the \R{CDFs for} both CL, FL, and closed-loop latency. \R{In practical applications, we can reduce the complexity of approximating $F_{\rm ET}(x)$ by controlling the desired approximation accuracy.} Based on Theorem 1, we conceive an algorithm for efficiently approximating $F_{\rm ET}(x)$ in practical applications. The details of this algorithm are shown in Alg. 1. Steps 6 and 9 are used for normalizing the PMF of $T^{\rm cm}_{\rm CD}+T^{\rm cm}_{\rm VI}$. Similar procedures can be used to compute $F_{\rm CL}(x)$ and $F_{{\rm T}_1}(x)$.

	\begin{figure*}[b]
		\normalsize \hrulefill
	\setcounter{equation}{52} 
		\begin{equation}\label{cgf3}
			M(s)= L(s)+\vartheta t_{\rm u}s+\frac{s^2}{2}\left(\frac{\epsilon_{\rm CD}N_{\rm CD}t_{\rm u}^2}{\left(1-\epsilon_{\rm CD}\right)^2}+\frac{\epsilon_{\rm VI}N_{\rm VI}t_{\rm u}^2}{\left(1-\epsilon_{\rm VI}\right)^2}\right), \, s<\min\left\{ \frac{\chi_{\rm RA}\alpha_{\rm RA}}{n_{\rm ET}\zeta_{\rm c}(\kappa)}, \frac{\chi_{\rm MEC}\alpha_{\rm MEC}\kappa}{n_{\rm ET}\zeta_{\rm d}(\kappa)},   \frac{\chi_{\rm MEC}\beta_{\rm VI}}{n_{\rm VI}}  \right\}.
		\end{equation}
	\end{figure*}
	
	\renewcommand{\algorithmicrequire}{\textbf{Input:}}  
	\renewcommand{\algorithmicensure}{\textbf{Output:}}  
	\begin{algorithm}[t]
		{
			\caption{Calculating CDF of $T_{\rm ET}$}
			\begin{algorithmic}[1]
				\REQUIRE
				parameter  $\Delta \in(0,1) $ for controlling the approximation accuracy
				\ENSURE
				$F_{\rm ET}(x)$
				\STATE Initialization: $k=N_{\rm CD}+N_{\rm VI}-1$, $\Theta=0$
				\REPEAT
				\STATE $k\leftarrow k+1$
				\STATE Calculate $p_k$ according to Eq. \eqref{lemma21}
				\UNTIL{$p_k<\Delta$}
				\STATE $
				\nu\leftarrow\mathbb{I}\{k>N_{\rm CD}+N_{\rm VI}\}\sum\limits_{i=N_{\rm CD}+N_{\rm VI}+1}^{k} p_i$
				\IF {$\nu>0$}
				\FORALL{$j=N_{\rm CD}+N_{\rm VI}+1, \cdots, k$}
				\STATE $p_j\leftarrow\left(1-p_{N_{\rm CD}+N_{\rm VI}}\right)\frac{p_j}{\nu}$
				\ENDFOR
				\ENDIF
				\IF {$\left\lfloor \frac{x}{t_{\rm u}} \right\rfloor < N_{\rm CD}+N_{\rm VI}$}
				\STATE $\Theta\leftarrow 0$
				\ELSE
				\FORALL{$l=N_{\rm CD}+N_{\rm VI}, \cdots, \min\left\{\left\lfloor \frac{x}{t_{\rm u}} \right\rfloor,k\right\}$}
				\STATE Calculate $F_1(x-lt_{\rm u})$ according to Eq. \eqref{lemma11}.
				\STATE $\Theta\leftarrow\Theta+p_lF_1(x-lt_{\rm u})$
				\ENDFOR
				\ENDIF
				\STATE $F_{\rm ET}(x)\leftarrow\Theta$
		\end{algorithmic}}
	\end{algorithm}

	\subsection{Continuous Approximation for a Large Number of Packets or Small Transmission Time of a Packet}

	In the last subsection, we present a general method to approximate the latency distribution. Since the CL, FL, and closed-loop latencies consist of sums of continuous and discrete random variables, we adopt both the \R{Lugannani-Rice (LR) formula \cite[Eq. (1.21)]{LR} and saddlepoint mass function \cite[Eq. (1.13)]{LR}} to obtain a satisfying approximation performance. However, in practical applications, the number of packets is usually large. Besides, the transmission time of a packet can be small in short-packet scenarios. In this subsection, we focus on these cases, \R{where either} the number of packets is large or $t_{\rm u}$ is small. \R{Under these conditions}, we can apply the central limit theorem (CLT) to approximate the sum of communication latency as normal random variables, \R{thereby treating $T_{\rm FL}$ and $T$ as} continuous random variables.
	
		We start with the analysis for the discrete components of $T_{\rm ET}$, which can be generalized to the analysis for the discrete components of $T$. \R{$T^{\rm cm}_{\rm CD}$ is the sum of $N_{\rm CD}$ i.i.d. discrete random variables, while $T^{\rm cm}_{\rm VI}$ is similar.  Under the condition that $N_{\rm CD}+N_{\rm VI}$ is large, the discrete random variable $T^{\rm cm}_{\rm CD}+T_{\rm VI}^{\rm cm}$ can be approximated by a normal random variable, which is different from the general analysis in Lemma 2. That is, $T^{\rm cm}_{\rm CD}+T^{\rm cm}_{\rm VI}\sim \mathcal{N}\left(\frac{N_{\rm CD}t_{\rm u}}{1-\epsilon_{\rm CD}}+\frac{N_{\rm VI}t_{\rm u}}{1-\epsilon_{\rm VI}},\frac{\epsilon_{\rm CD}N_{\rm CD}t_{\rm u}^2}{\left(1-\epsilon_{\rm CD}\right)^2}+\frac{\epsilon_{\rm VI}N_{\rm VI}t_{\rm u}^2}{\left(1-\epsilon_{\rm VI}\right)^2}\right)$.} As analyzed in \cite{suman2023}, we are interested in scenarios in which the negative value from Gaussian approximation has a negligible effect on the analysis. 
	Moreover, $T_{\rm CD}^{\rm cm}$ and $T_{\rm VI}^{\rm cm}$ are both $t_{\rm u}$-lattice random variables. Thus,  we can treat $T_{\rm CD}^{\rm cm}$ and $T_{\rm VI}^{\rm cm}$ as continuous random variables when $t_{\rm u}$ is small. There are two possibilities \R{for scenarios with small $t_{\rm u}$} from the perspective of the number of packets. \R{The first possibility} is that $N_{\rm CD}+N_{\rm VI}$ is small. Since $t_{\rm u}$ is also small, the corresponding transmission time has a small influence on the distribution of $T$. The second possibility is that the $N_{\rm CD}+N_{\rm VI}$ is large. Then we treat $T^{\rm cm}_{\rm CD}+T_{\rm CD}^{\rm cm}$ as a continuous random variable by CLT. In summary, we can use CLT to approximate the distribution of $T_{\rm CD}^{\rm cm}+T_{\rm VI}^{\rm cm}$ with large $N_{\rm CD}+N_{\rm VI}$ or small $t_{\rm u}$.
	
	Based on this continuous approximation, we can directly obtain the CDF and PDF approximation of $T_{\rm ET}$ according to the LR formula.  We define the following constants as 
	\setcounter{equation}{47} 
	\begin{equation}
		\Psi=\theta+\vartheta t_{\rm u},
	\end{equation}
	where 
	\begin{equation}\label{expectmds}
		\begin{aligned}
			\vartheta=\frac{N_{\rm CD}}{1-\epsilon_{\rm CD}}+\frac{N_{\rm VI}}{1-\epsilon_{\rm VI}}.
		\end{aligned}
	\end{equation}
	We also define that
	\begin{equation}
		\Upsilon=\left(\iota_1^{\frac{2}{3}}+\frac{\epsilon_{\rm CD}N_{\rm CD}t_{\rm u}^2}{\left(1-\epsilon_{\rm CD}\right)^2}+\frac{\epsilon_{\rm VI}N_{\rm VI}t_{\rm u}^2}{\left(1-\epsilon_{\rm VI}\right)^2}\right)^{\frac{3}{2}}.
	\end{equation}

	\textbf{Lemma 3}. Under the condition that $N_{\rm CD}+N_{\rm VI}$ is large or $t_{\rm u}$ is small,
	\begin{itemize}
		\item the CDF of $T_{\rm ET}$ is approximated by
		\begin{equation}\label{coro11}
			\!\!\! \begin{aligned}
				\R{\hat{F}_{\rm ET}(x)\approx\begin{cases}
						\Phi(r_x)+\phi(r_x)\left ( \frac{1}{r_x}-\frac{1}{z_x} \right),  \quad &x\neq \Psi,\\
						\frac{1}{2}+\frac{\iota_2}{6\sqrt{2\pi }\Upsilon}, \quad & x=\Psi,
				\end{cases}}
			\end{aligned}
		\end{equation}
		where
		\begin{equation}
			\!\!\!\begin{aligned}
				r_x&={\rm sign}\left(\Xi_3\left(x\right)\right)\sqrt{2x\Xi_3\left(x\right)-2M(\Xi_3(x))},\,  \text{and} \\
				z_x&=\Xi_3(x)\sqrt{M''(\Xi_3(x))}.
			\end{aligned}
		\end{equation}
		$M(s)$ is given in Eq. \eqref{cgf3}. $\Xi_3(\cdot)$ is the inverse function of $M'(s)$.
		
		\item the PDF of $T_{\rm ET}$ is approximated by
		\setcounter{equation}{53}
		\begin{equation}\label{pdfconti}
			\begin{aligned}
				\R{\hat{f}_{\rm ET}(x)\approx\frac{\exp\left(M\left(\Xi_3(x)\right)-x\Xi_3(x)\right)}{\sqrt{2\pi M''(\Xi_3(x))}}, x>0.}
			\end{aligned}
		\end{equation}
	\end{itemize}

	\begin{IEEEproof}
		The proof is similar to the proof of Lemma 1. The CGF is shown in Eq. \eqref{cgf3}. $\Psi$ is the expectation of $T_{\rm ET}$. It is obvious that $M'(s)$ is an increasing function of $s$. Thus, the inverse function of $M'(s)$ exists, which we denote by $\Xi_3(\cdot)$. Besides, we find that $M''(0)^{3/2}=\Upsilon$ and $M'''(0)=\iota_2$. \R{Thus, we obtain Eq. \eqref{coro11} according to LR formula \cite[Eq. (1.21)]{LR}. With the help of the saddlepoint density function [Eqs. (1.4) and (1.5)]\cite{LR}, we obtain Eq. \eqref{pdfconti}.}
	\end{IEEEproof}

	By combining CLT with SPA, we can directly obtain the CDF of $T_{\rm ET}$. The same method can be applied to obtaining the CDF of $T_{1}$ \R{under} the condition that $N_{\rm HL}+N_{\rm LL}+N_{\rm CD}+N_{\rm VI}$ \R{is} large or $t_{\rm u}$ is small. Let $\tilde{M}(s)$ denote the CGF of $T_1$, which is \R{provided} in Eq. \eqref{tms}, where $s<\min\left\{  \frac{\chi_{\rm MEC}\beta_{\rm LL}}{N_{\rm HL}n_{\rm HL}} , \frac{\chi_{\rm RA}\alpha_{\rm RA}}{n_{\rm ET}\zeta_{\rm c}(\kappa)}, \frac{\chi_{\rm MEC}\alpha_{\rm MEC}\kappa}{n_{\rm ET}\zeta_{\rm d}(\kappa)},   \frac{\chi_{\rm MEC}\beta_{\rm VI}}{n_{\rm ET}}   \right\}$. Besides, we also define $\tilde{\Psi}$ and $\tilde{\Upsilon}$ as follows
	\setcounter{equation}{55}
	\begin{equation}
		\!\tilde{\Psi}=\tilde{\theta}+\left(\vartheta+\frac{N_{\rm HL}}{1-\epsilon_{\rm HL}}+\frac{N_{\rm LL}}{1-\epsilon_{\rm LL}}\right)t_{\rm u}.
	\end{equation} 
	\begin{equation}
		\!\!\!\!\tilde{\Upsilon}\!\!=\!\!\left(\!\Upsilon^{\frac{2}{3}}\!+\!\!\frac{N_{\rm HL}^2n_{\rm HL}^2\alpha_{\rm MEC}}{\chi^2_{\rm MEC}\beta_{\rm LL}^2}\!+\!\frac{\epsilon_{\rm CD}N_{\rm CD}t_{\rm u}^2}{\left(1-\epsilon_{\rm CD}\right)^2}\!+\!\frac{\epsilon_{\rm VI}N_{\rm VI}t_{\rm u}^2}{\left(1-\epsilon_{\rm VI}\right)^2}\!\right)^{\frac{3}{2}}\!\!.
	\end{equation}
	
	\begin{figure*}[b]
		\normalsize \hrulefill
		\setcounter{equation}{54} 
			\begin{equation}\label{tms}
			\tilde{M}(s)=M(s)-\alpha_{\rm MEC}\ln\left(1-\frac{N_{\rm HL}n_{\rm HL}s}{\chi_{\rm MEC}\beta_{\rm LL}}\right)+\left(\frac{N_{\rm HL}t_{\rm u}}{1-\epsilon_{\rm HL}}+\frac{N_{\rm LL}t_{\rm u}}{1-\epsilon_{\rm LL}}\right)s+\frac{s^2}{2}\left(\frac{\epsilon_{\rm HL}N_{\rm HL}t_{\rm u}^2}{\left(1-\epsilon_{\rm HL}\right)^2}+\frac{\epsilon_{\rm LL}N_{\rm LL}t_{\rm u}^2}{\left(1-\epsilon_{\rm LL}\right)^2}\!\right).
		\end{equation}
	\setcounter{equation}{57}
		\begin{equation}\label{theo21}
			\begin{aligned}
				\mathbb{E}\left\{\left(T_{\rm ET}-\tau_{\rm PF}\right)^+\right\}\approx\left(\theta+\vartheta t_{\rm u}-\tau_{\rm PF}\right)\left(1-\hat{F}_{\rm ET}(\tau_{\rm PF})\right)+\frac{\tau_{\rm PF}-\theta-\vartheta t_{\rm u}}{\Xi_3(\tau_{\rm PF})}\hat{f}_{\rm ET}(\tau_{\rm PF}), \quad \text{if $\tau_{\rm PF}\neq \theta+\vartheta t_{\rm u}$}.
			\end{aligned}
		\end{equation}
	\end{figure*}
	
	By substituting $\tilde{M}(s)$, $\tilde{\iota}_2$ (defined below Eq. \eqref{tt1}),  $\tilde{\Psi}$, and $\tilde{\Upsilon}$ into Eqs. \eqref{coro11} and \eqref{pdfconti}, we can obtain $\hat{F}_{{\rm T}_1}(x)$ and $\hat{f}_{{\rm T}_1}(x)$, where $\hat{F}_{{\rm T}_1}(x)$ and $\hat{f}_{{\rm T}_1}(x)$ are the continuous approximation for the CDF and PDF of $T_1$, respectively. We then present the continuous approximation for the CDF of $T$ in Theorem 2.

	\textbf{Theorem 2}. Under the condition that $N_{\rm HL}+N_{\rm LL}+N_{\rm CD}+N_{\rm VI}$ is large or $t_{\rm u}$ is small, the CDF of the closed-loop latency is  approximated by
	\setcounter{equation}{57} 
	\begin{equation}\label{th2mm}
		\R{\hat{F}_{\rm T}(x)\approx\mathbb{I}\{x>\tau_{\rm PF}\}F_{\rm CL}(x-\tau_{\rm PF})\hat{F}_{{\rm T}_1}(x).}
	\end{equation}

	Note that if $N_{\rm LL}+N_{\rm HL}$ is large or $t_{\rm u}$ is small, we can also use the continuous approximation $\hat{F}_{\rm CL}(x)$ to \R{substitute} $F_{\rm CL}(x)$ in Eq. \eqref{th2mm}. However, $N_{\rm LL}$ is usually equal to 1 in practical scenarios.  \R{Besides, $T_{\rm CL}$ consists of only three latency components, among which the CDF of $T_{\rm LL}^{\rm cp}$ is known.} Thus, we \R{employ} the general approximation for the CDF of $T_{\rm CL}$ in Eq. \eqref{th2mm} to ensure high accuracy while \R{maintaining} efficiency.

	\subsection{Efficient Evaluation for Different Kinds of Constraints and Properties for Optimization}
	
	In this subsection, we will propose the methods for efficiently evaluating some kinds of constraints with the help of conclusions in Sections IV-A and IV-B. Then, we will present some properties of the communication and computation processes in this system, which are useful for further optimization.

In Sections IV-A and IV-B, we have derived the CDFs of $T_{\rm CL}$, $T_{\rm ET}$, $T_{\rm FL}$, and $T$. Besides, the expectation and variance of each latency component are known. Thus, all constraints mentioned in Section II-B can be characterized by these conclusions. However, \R{certain} constraints can be evaluated more efficiently, \R{prompting us to present the efficient evaluating methods based on the SPA-based methods}.

	\R{Eq. \eqref{asf} imposes a constraint on the conditional expectation $\mathbb{E}\left\{\left(T_{\rm ET}-\tau_{\rm PF}\right)^+\right\}$. An intuitive approach to calculate this would be to use the PDF of $T_{\rm ET}$.} However, we can not obtain the PDF of $T_{\rm ET}$ for general cases since $T_{\rm ET}$ may not be a continuous random variable. \R{While it appears feasible to derive $f_{\rm ET}(x)$ by Eq. \eqref{pdfconti} under the assumption that $N_{\rm CD}+N_{\rm VI}$ is large or $t_{\rm u}$ is small, the closed-form expression of $\Xi_3(x)$ in Eq. \eqref{pdfconti} is hard to obtain. This complication makes the computation of $f_{\rm ET}(x)$ over a continuous interval impractical.} One \R{potential} method to approximate the PDF of $T_{\rm ET}$ is \R{through discretization, as used} in our previous work \cite{icc1}.

	Though we can design a discrete approximation for the PDF of $T_{\rm ET}$ based on Lemma 3, it is \R{computationally intensive due to the need to solve SPA equations at numerous points when granularity is small.} Therefore, we present an efficient method to derive $\mathbb{E}\left\{\left(T_{\rm ET}-\tau_{\rm PF}\right)^+\right\}$ in Corollary 1.

	\textbf{Corollary 1}. Under the condition that $N_{\rm CD}+N_{\rm VI}$ is large or $t_{\rm u}$ is small,   $	\mathbb{E}\left\{\left(T_{\rm ET}-\tau_{\rm PF}\right)^+\right\}$ can be approximated by Eq. \eqref{theo21}, where $\hat{F}_{\rm ET}(\tau_{\rm PF})$, $\hat{f}_{\rm ET}(\tau_{\rm PF})$, and $\Xi_3(\tau_{\rm PF})$ are given in Lemma 3.
	
	\begin{IEEEproof}
		Corollary 1 is based on the Eq. (2) in \cite{cond}. \R{Note that Eq. (2) in \cite{cond} requires that the random variable should be continuous. As we analyzed before Lemma 3, $T_{\rm CD}^{\rm cm}+T_{\rm VI}^{\rm cm}$ can be approximated by a normal random variable by CLT under the condition that $N_{\rm CD}+N_{\rm VI}$ is large or $t_{\rm u}$ is small. By this means, $T_{\rm ET}$ can be treated as a continuous random variable with CDF $\hat{F}_{\rm ET}(\tau_{\rm PF})$ and  $\hat{f}_{\rm ET}(\tau_{\rm PF})$. }
	\end{IEEEproof}

Based on Eq. \eqref{theo21} in Corollary 1, we can evaluate $\mathbb{E}\left\{\left(T_{\rm ET}-\tau_{\rm PF}\right)^+\right\}$ by only calculating the CDF and PDF of $T_{\rm ET}$ \R{at a single point}, which is more efficient than the discrete approximation method.

	Then, we turn our focus to Eq. \eqref{v3}. The communication latency of PF data is stochastic for this constraint. To calculate $\Pr\{|T_{\rm ET}-T_{\rm PF}|<\tau_{\rm td}\}$, it is equivalent to calculate
	\setcounter{equation}{58}
	\begin{equation}
		\Pr\{T_{\rm ET}-T_{\rm PF}<\tau_{\rm td}\}-\Pr\{T_{\rm ET}-T_{\rm PF}<-\tau_{\rm td}\}.
	\end{equation}
	Thus, we can \R{employ} a similar method proposed in Theorem 1 to adopt the SPA-based method by first deriving the CGF of $T_{\rm ET}-T_{\rm PF}$. Due to space limits, we omit the details.

	Then, we provide some useful properties for optimizing the system's performance. For the PF data, we present a conclusion on the period allocation for the case with CSIRT and FSD to minimize the average power consumption for the PF data transmission. For the ET data, we offer a conclusion related to the choice of compression ratio. \R{For convenience, we let}
	\begin{equation}
		\!\!\!   \R{\varphi_j   \!  = \!   \left[\frac{N_0B_{\rm PF}}{c_1d_{j}^{-\ell_{j}}}\left(2^{\frac{n_{\rm PF}}{T_{\rm s}B_{\rm PF}}}-1\right)\right]^{-\frac{1}{c_3+1}}, \, j\in\{{\rm PF1 },{\rm PF2}\}.}
	\end{equation} 
	
	\textbf{Lemma 4}. \R{Given that the transmission of PF data with CSIRT and FSD, the period allocation to achieve a minimum $P_{\rm PF1}+P_{\rm PF2}$ is $\tau_{\rm PF1}=\frac{\tau_{\rm PF}\varphi_{\rm PF2}}{\varphi_{\rm PF1}+\varphi_{\rm PF2}}$ and $\tau_{\rm PF2}=\frac{\tau_{\rm PF}\varphi_{\rm PF1}}{\varphi_{\rm PF1}+\varphi_{\rm PF2}}$.}
	\begin{IEEEproof}
	 See Appendix C.
	\end{IEEEproof}

	\textbf{Lemma 5}. \R{For $\zeta_{\rm d}(\kappa)$ satisfying Eq. \eqref{decom1} with $\psi\geq 1$ and \eqref{decom2} with $\omega_7>1+\frac{\omega_8}{\omega_6}$,} $
	\frac{\kappa}{\zeta_{\rm d}({\kappa})}$ is a decreasing function of $\kappa$. For $\zeta_{\rm d}(\kappa)$ shown in Eq. \eqref{decom2} with $\omega_7\leq 1$, $
	\frac{\kappa}{\zeta_{\rm d}({\kappa})}$ is an increasing function of $\kappa$.
	\begin{IEEEproof}
		See Appendix D.
	\end{IEEEproof}
	
	In Lemma 4, we provide a closed-form resource allocation scheme for the transmission of PF data. In Lemma 5, we present conclusions \R{regarding} the monotonic properties of $\frac{\kappa}{\zeta_{\rm d}(\kappa)}$. \R{These conclusions offer guidance for the optimization in the next subsection.}

	\subsection{\R{A Specific Optimization Problem}}
	
	\R{In this subsection, we will show a specific case of Problem \eqref{optpro} and demonstrate how to solve this problem with the help of the proposed theorems and lemmas.}
	
	\R{For simplicity, we consider optimizing the compression ratio $\kappa$ and transmission power $P_{\rm PF1}$, $P_{\rm PF2}$, and $P_{\rm CD}$, while treating the bandwidth and other transmission powers as fixed. By adopting the constraints from vertex 8 in Fig. \ref{constraints}, the optimization problem is formulated as }
	\begin{subequations}\label{opj}
		\R{\begin{align}
				\min_{\kappa, P_{\rm PF1}, P_{\rm PF2},  P_{\rm CD}} & \varepsilon \Pr\left\{T>T_{\rm th}\right\}\!+\!\frac{1-\varepsilon}{P_{\rm max}}\left(P_{\rm PF1}\!+\!P_{\rm PF2}\!+\!P_{\rm CD}\right) \nonumber \\
				\mbox{s.t.}\quad
				&\tau_{\rm PF1}+\tau_{\rm PF2}\leq \tau_{\rm PF}, \\
				&\Pr\{T_{\rm ET}<\tau_{\rm PF}\}\geq \eta_{\rm ts}, \label{opj2}\\
				&{\rm Var}\{T_{\rm ET}-\tau_{\rm PF}\}\leq \rho_{\rm ts},\label{varcon}\\
				& R_{\rm CD}=\frac{t_u n_{\rm CD}^{(1)}}{\kappa}, \label{powercd} \\
				& \zeta_{\rm c}(\kappa) =\exp\left(\psi \kappa\right)-\exp\left(\psi\right), \\
				& \zeta_{\rm d}(\kappa)= \omega_0\left(\exp\left(\psi \kappa\right)-\exp\left(\psi\right)\right),
		\end{align}}
	\end{subequations}
	\R{where $\psi\geq 1$. $n_{\rm CD}^{(1)}$ represents the packet size of the compressed data when $\kappa=1$. Eq. \eqref{powercd} means that we change the size of a compressed-data packet to fix $N_{\rm CD}$. In this problem, $d(T)=\Pr\{T>T_{\rm th}\}$ represents the aim to reduce the delay violation probability with threshold $T_{\rm th}$ for the closed-loop latency. $P_{\rm max}\geq P_{\rm PF1}+P_{\rm PF2}+P_{\rm CD}$ is used to normalize the power-related term in the optimization objective.  }

	\R{Then, we provide the analysis for this optimization problem. According to Eq. \eqref{equicon}, Eq. \eqref{varcon} is equivalent to  }
	\begin{equation}\label{varcon1}
		\R{\xi^{\rm c}_{\rm ET}+\xi_{\rm ET}^{\rm d}+\xi^{\rm cp}_{\rm VI}+\frac{\epsilon_{\rm CD}N_{\rm CD}t_{\rm u}^2}{\left(1-\epsilon_{\rm CD}\right)^2}+\frac{\epsilon_{\rm VI}N_{\rm VI}t_{\rm u}^2}{\left(1-\epsilon_{\rm VI}\right)^2}\leq \rho_{\rm ts},}
	\end{equation}
	\R{where $\xi^{\rm c}_{\rm ET}$, $\xi_{\rm ET}^{\rm d}$, and $\xi^{\rm cp}_{\rm VI}$ are defined in Section III. Both $\xi^{\rm c}_{\rm ET}$ and $\xi_{\rm ET}^{\rm d}$ are functions of $\kappa$. According to Lemma 5, $\xi^{\rm c}_{\rm ET}+\xi_{\rm ET}^{\rm d}$ is an increasing function of $\kappa$ with $\psi\geq 1$. Thus, we can determine an upper bound of $\kappa$ through Eq. \eqref{varcon1}, which we denote as $\hat{\kappa}_1$. $\hat{\kappa}_1$ is the solution of the following equation, which can be solved efficiently through binary search.}
	\begin{equation}\label{varcon2}
		\R{\xi^{\rm c}_{\rm ET}+\xi_{\rm ET}^{\rm d}= \rho_{\rm ts}-\xi^{\rm cp}_{\rm VI}-\frac{\epsilon_{\rm CD}N_{\rm CD}t_{\rm u}^2}{\left(1-\epsilon_{\rm CD}\right)^2}-\frac{\epsilon_{\rm VI}N_{\rm VI}t_{\rm u}^2}{\left(1-\epsilon_{\rm VI}\right)^2}.}
	\end{equation}

	\R{Given $\tau_{\rm PF}$, the optimal choice of $\tau_{\rm PF1}$ and $\tau_{\rm PF2}$ for minimizing $P_{\rm PF1}+P_{\rm PF2}$ are determined by Lemma 4. By substituting $\tau_{\rm PF1}=\frac{\tau_{\rm PF}\varphi_{\rm PF2}}{\varphi_{\rm PF1}+\varphi_{\rm PF2}}$ and $\tau_{\rm PF2}=\frac{\tau_{\rm PF}\varphi_{\rm PF1}}{\varphi_{\rm PF1}+\varphi_{\rm PF2}}$ into Eq. \eqref{dcc}, we have}
	\begin{equation}\label{ppf}
		\R{\begin{aligned}
				P_{\rm PF1}+P_{\rm PF2} &=c_2\left(\varphi_{\rm PF1}^{-c_3-1}+\varphi_{\rm PF2}^{-c_3-1}\right)+\\
				&\left(\frac{\tau_{\rm PF}\varphi_{\rm PF1}\varphi_{\rm PF2}}{\varphi_{\rm PF1}+\varphi_{\rm PF2}}\right)^{-c_3}\left(\varphi_{\rm PF1}^{-1}+\varphi_{\rm PF2}^{-1}\right).
		\end{aligned}}
	\end{equation}
	\R{For $P_{\rm CD}$, we can express it as a function of $\kappa$. Based on Eqs. \eqref{orate} and \eqref{powercd}, we have}
	\begin{equation}\label{p3}
		\R{P_{\rm CD}=\frac{-N_0B_{\rm CD}}{d_{\rm RB}^{-\ell_{\rm CD}}\ln(1-\epsilon_{\rm CD})}\left(2^{\frac{t_{\rm u}n_{\rm CD}^{(1)}}{B_{\rm CD}}\cdot\frac{1}{\kappa}}-1\right)}.
	\end{equation}
	
	\R{For the objective function and Eq. \eqref{opj2}, we adopt the SPA approximation in Theorem 1. We start our analysis with Eq. \eqref{opj2}. Since the distribution of transmission time is fixed in this problem, our focus is on the computation, compression, and decompression time. In this case, the value of $\kappa$ will influence $\Pr\{T_{\rm ET}<\tau_{\rm PF}\}$ through changing the distribution of $T^{\rm c}_{\rm ET}+T^{\rm d}_{\rm ET}+T^{\rm cp}_{\rm VI}$. Since in Lemma 5 we have proved that $\frac{\kappa}{\zeta_{\rm d}(\kappa)}$ is a decreasing function of $\kappa$ with $\Psi\geq 1$, the CDF of $T^{\rm c}_{\rm ET}+T^{\rm d}_{\rm ET}+T^{\rm cp}_{\rm VI}$ is a decreasing function of $\kappa$. Thus, the SPA $F_1(\tau_{\rm PF})$ can be seen as a decreasing function of $\kappa$. Therefore, we can obtain another upper bound of $\kappa$ through Eq. \eqref{opj2}, which we denote by $\hat{\kappa}_2$. Combining $\hat{\kappa}_2$ with $\hat{\kappa}_1$ and $\kappa_{\rm max}$, the upper bound of $\kappa$ is given by $\kappa\leq \min\left\{\kappa_{\rm max}, \hat{\kappa}_1, \hat{\kappa}_2\right\}$.}
	
	\R{Similarly, $\Pr\{T>T_{\rm th}\}$ is an increasing function of $\kappa$. Therefore, there is a tradeoff between $\Pr\{T>T_{\rm th}\}$ and $P_{\rm PF1}+P_{\rm PF2}+P_{\rm CD}$ since $P_{\rm CD}$ is a decreasing function of $\kappa$ as shown in Eq. \eqref{p3}. In contrast to the grid search used in \cite{suman2023}, we can utilize the gradient-based method to find the near-optimal solution to Problem \eqref{opj} thanks to the analytical expressions obtained from SPA. Since the convexity of $\Pr\{T>T_{\rm th}\}$ is unknown, we can adopt the gradient-descent method with random start points to relieve the probability of obtaining the local optimal solutions. It is an important future work to consider more advanced algorithms for solving this problem, which is beyond the scope of this paper. Besides, through the numerical results shown in Section V, we find that the objective function of Problem \eqref{opj} is a unimodal function of $\kappa$. Therefore, we can use ternary search \cite{ternary} to find the optimal solution. Let $N$ denote the number of feasible searching points. The computational complexity of the ternary search is $O(\log_3 N)$ \cite{ternary}, which is lower than that of the grid search. }
	
	\R{We can use the numerical gradient to approximately obtain the gradient information. Moreover, we can also try to characterize the analytical gradient since we have the analytical expression of the latency distribution. To use the gradient-based method, we need to show the existence of $\frac{ \partial\Pr\{T>T_{\rm th}\}}{\partial\kappa}$. According to Theorem 1, it is equivalent to show the existence of $\frac{\partial\tilde{F}_1(x)}{\partial\kappa}$. We divide the following discussion into three parts. We first discuss $\frac{\partial s^{\star}}{\partial \kappa}$, where $s^{\star}$ is the solution to the saddlepoint function of $\tilde{F}_1(x)$. The saddlepoint equation of $\tilde{F}_1(x)$ is given by $\tilde{L}'(s^{\star})=x$, where $\tilde{L}'(\cdot)$ is also a function of $\kappa$. By adopting implicit differential, we can obtain $\frac{\partial s^{\star}}{\partial \kappa}$ from the saddlepoint equation. For the simplicity of expression, we let $s^{\star}=Z(\kappa)$ denote the solution of the saddlepoint equation. 
		With abuse of notations, we denote the CGF by $\tilde{L}(\kappa,s^{\star})$ in the following discussion. Then, we discuss $\frac{{\rm d} u_{x}}{{\rm d} \kappa}$. For the simplicity of expression, we let $\Omega_0(\kappa)=\tilde{L}(\kappa,s^{\star})|_{s^{\star}=Z(\kappa)}$,  $\Omega_1(\kappa)=\frac{\partial\tilde{L}(\kappa,s^{\star})}{\partial s^{\star}}\big|_{s^{\star}=Z(\kappa)}$ and $\Omega_2(\kappa)=\frac{\partial^2\tilde{L}(\kappa,s^{\star})}{\partial (s^{\star})^2}\big|_{s^{\star}=Z(\kappa)}$. According to Eq. \eqref{uv}, we obtain that}
	\begin{equation}
		\R{\frac{{\rm d} u_{x}}{{\rm d} \kappa}=\sqrt{\Omega_2(\kappa)}\cdot\frac{{\rm d} Z(\kappa)}{{\rm d} \kappa}+\frac{Z(\kappa)}{2\sqrt{\Omega_2(\kappa)}}\cdot\frac{{\rm d} \Omega_2(\kappa)}{{\rm d} \kappa}   }.
	\end{equation}   
	\R{Finally, we discuss $\frac{{\rm d} v_x}{{\rm d} \kappa}$. According to Eq. \eqref{uv}, we have}
	\begin{equation}\label{add6}
		\R{v_x{\rm d} v_x=\left(x\frac{{\rm d}Z(\kappa)}{{\rm d}\kappa}-\frac{{\rm d}\Omega_0(\kappa)}{{\rm d}\kappa}\right){\rm d}\kappa}.
	\end{equation}
	\R{Based on Eq. \eqref{add6}, we have}
	\begin{equation}\label{add7}
		\R{\frac{{\rm d}v_x}{{\rm d}\kappa}=\frac{x\frac{{\rm d}Z(\kappa)}{{\rm d}\kappa}-\frac{{\rm d}\Omega_0(\kappa)}{{\rm d}\kappa}}{{\rm sign}\left(Z(\kappa)\right)\sqrt{2xZ(\kappa)-2\Omega_0(\kappa)}}.}
	\end{equation}
	\begin{figure*}[t]
	\begin{minipage}[b]{0.24\linewidth}
		\centering
		\subfigure[CDF of $T_{\rm ET}$.]{\includegraphics[width=4.8cm]{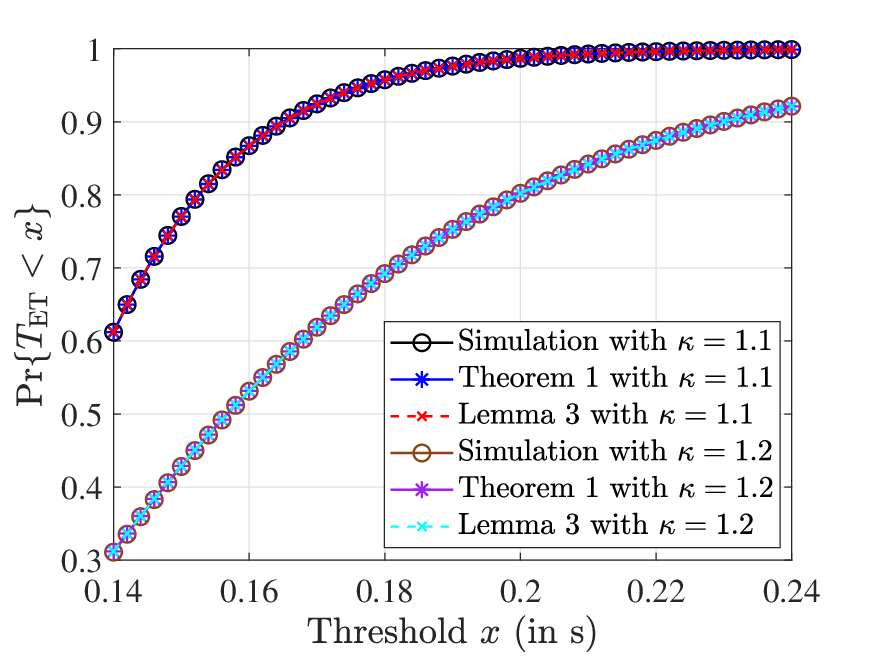}}
		\subfigure[Approximation error of  $T_{\rm ET}$.]{\includegraphics[width=4.8cm]{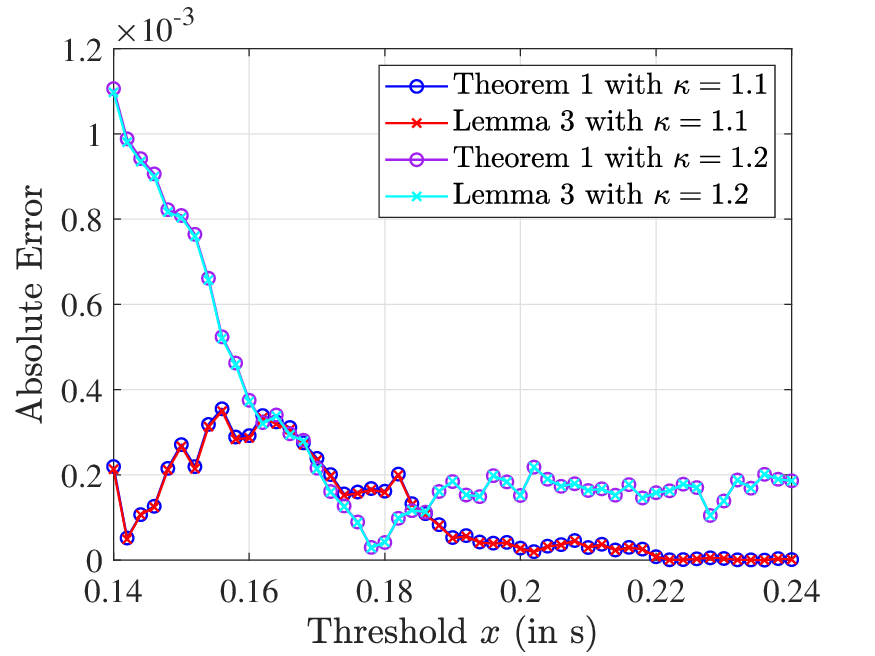}}
		\subfigure[CDF of $T$.]{\includegraphics[width=4.8cm]{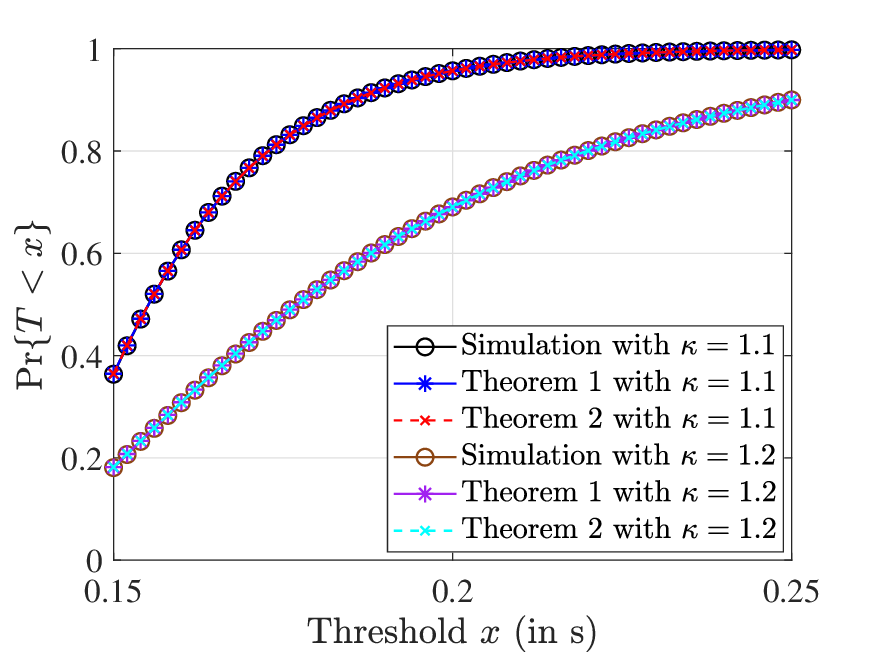}}
		\subfigure[\R{Approximation error of  $T$}]{\includegraphics[width=4.8cm]{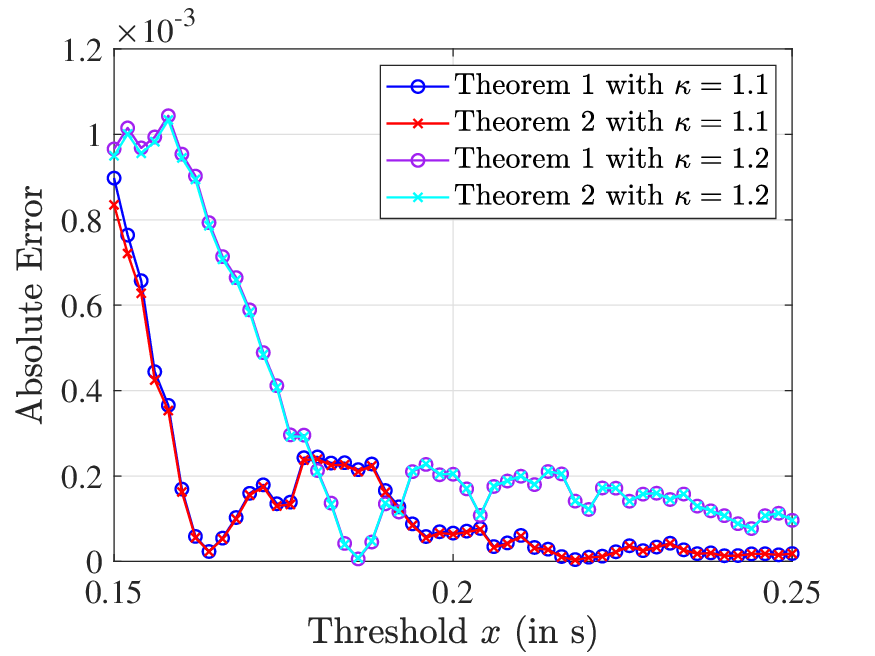}}
		\caption{$T_{\rm ET}$ and $T$ with $t_{\rm u}=5$ ms, $N_{\rm CD}^{(1.1)}=20$, $N_{\rm VI}=3$, $\epsilon_{\rm CD}=10^{-3}$, and $\epsilon_{\rm VI}=10^{-4}$.}
		\label{fig4}
	\end{minipage}
	\hspace{0.6mm}
	\begin{minipage}[b]{0.24\linewidth}
		\centering
		\subfigure[CDF of $T_{\rm ET}$.]{\includegraphics[width=4.8cm]{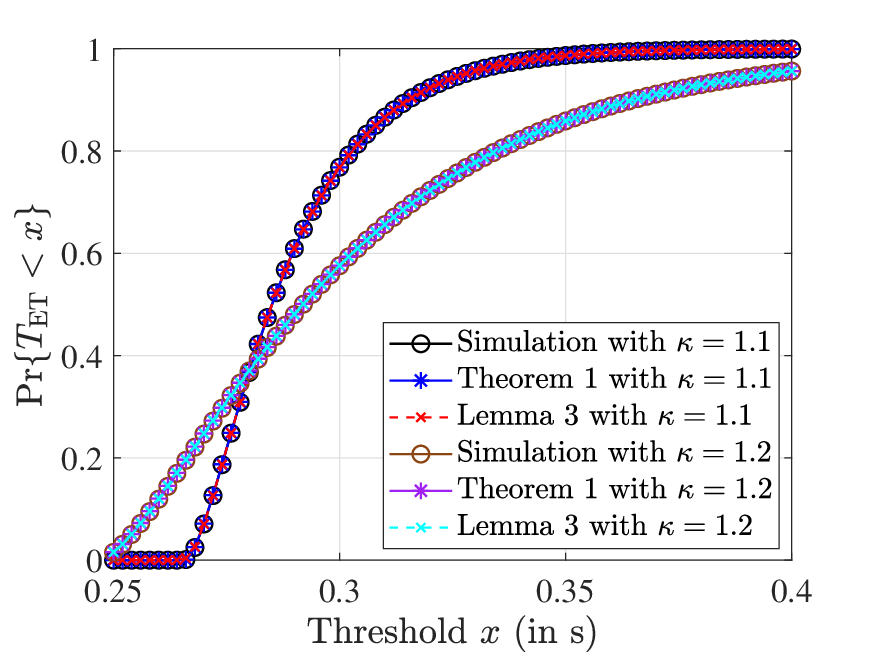}}
		\subfigure[Approximation error of $T_{\rm ET}$.]{\includegraphics[width=4.8cm]{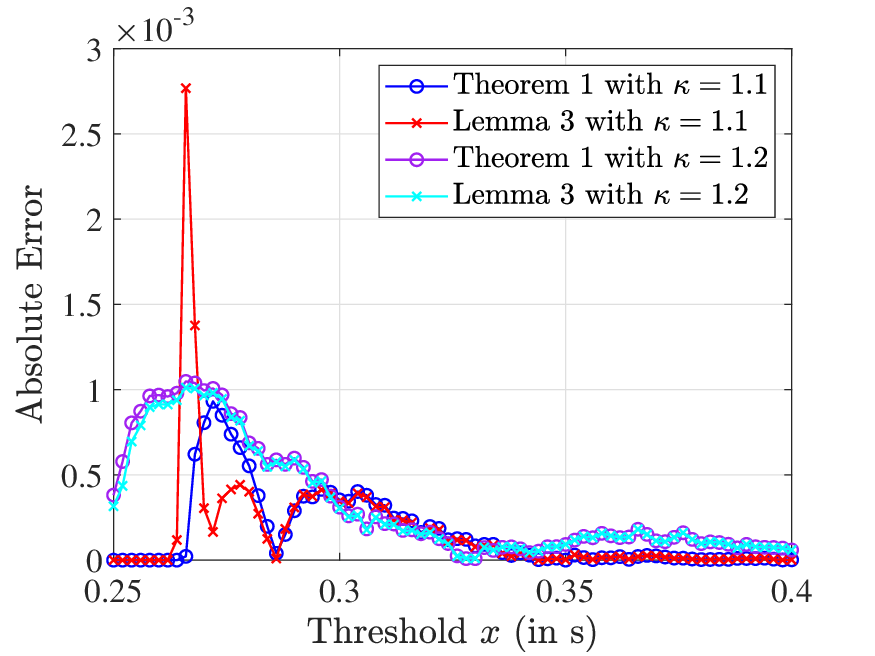}}
		\subfigure[CDF of $T$.]{\includegraphics[width=4.8cm]{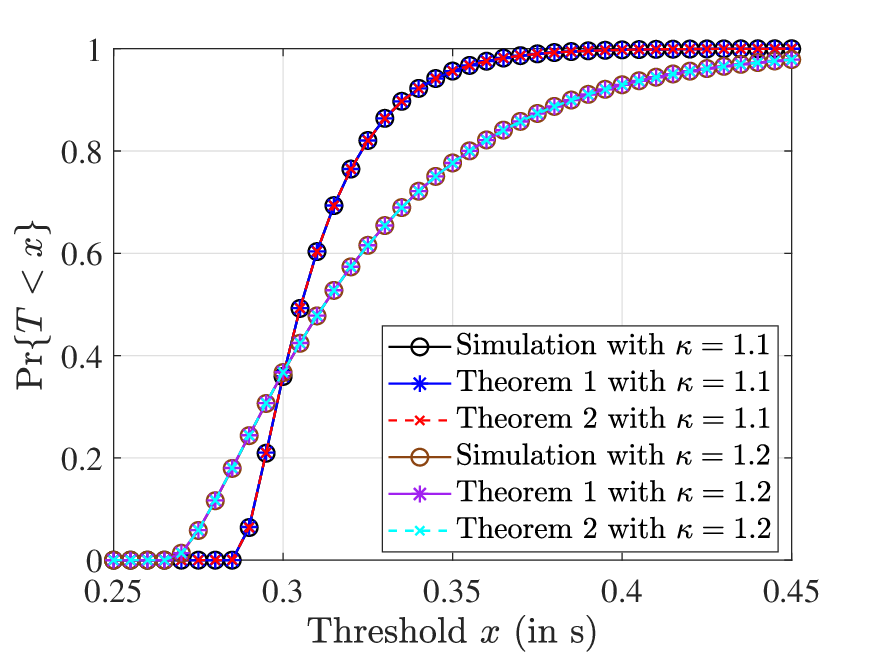}}
		\subfigure[\R{Approximation error of  $T$}]{\includegraphics[width=4.8cm]{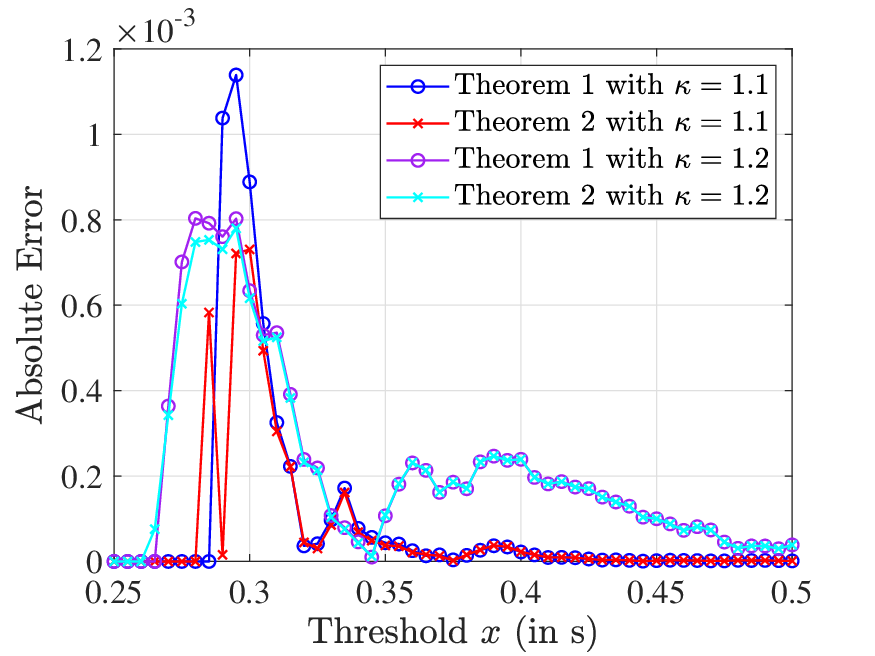}}
		\caption{$T_{\rm ET}$ and $T$ with $t_{\rm u}=5$ ms, $N_{\rm CD}^{(1.1)}=50$, $N_{\rm VI}=3$, $\epsilon_{\rm CD}=10^{-3}$, and $\epsilon_{\rm VI}=10^{-4}$.}
		\label{fig5}
	\end{minipage}
	\hspace{0.6mm}
	\begin{minipage}[b]{0.24\linewidth}
		\centering
		\subfigure[CDF of $T_{\rm ET}$.]{\includegraphics[width=4.8cm]{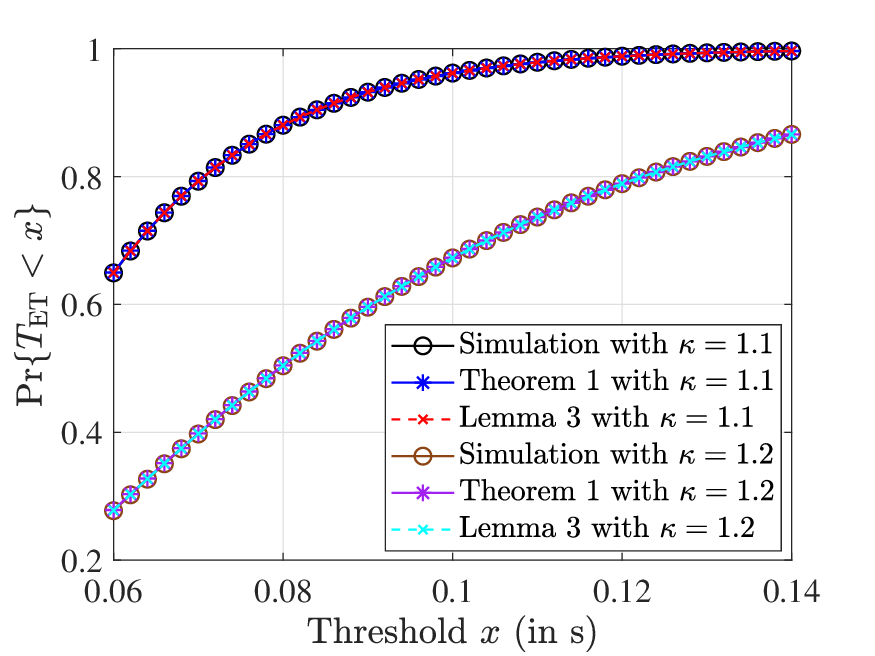}}
		\subfigure[Approximation error of $T_{\rm ET}$.]{\includegraphics[width=5cm]{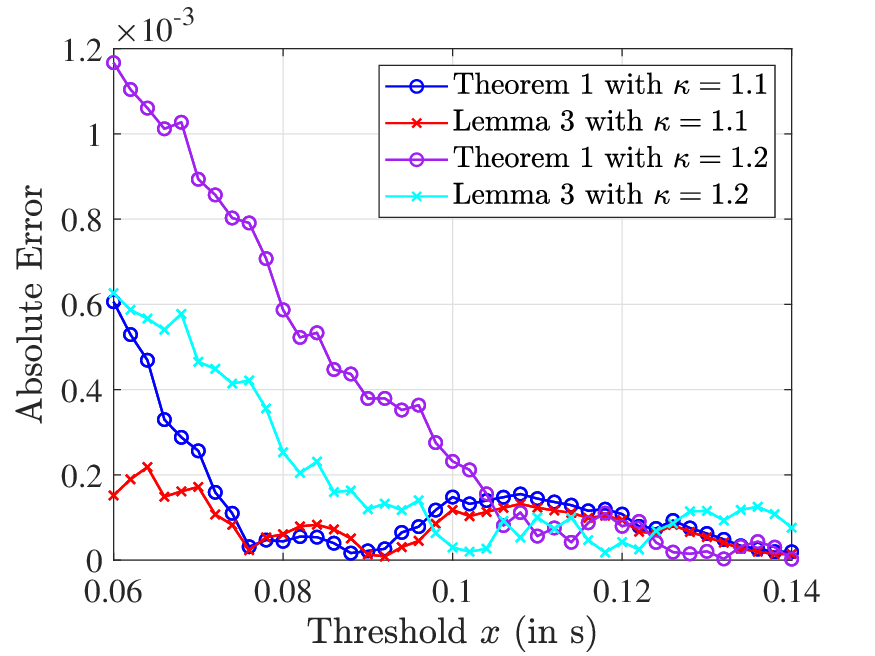}}
		\subfigure[CDF of $T$.]{\includegraphics[width=4.8cm]{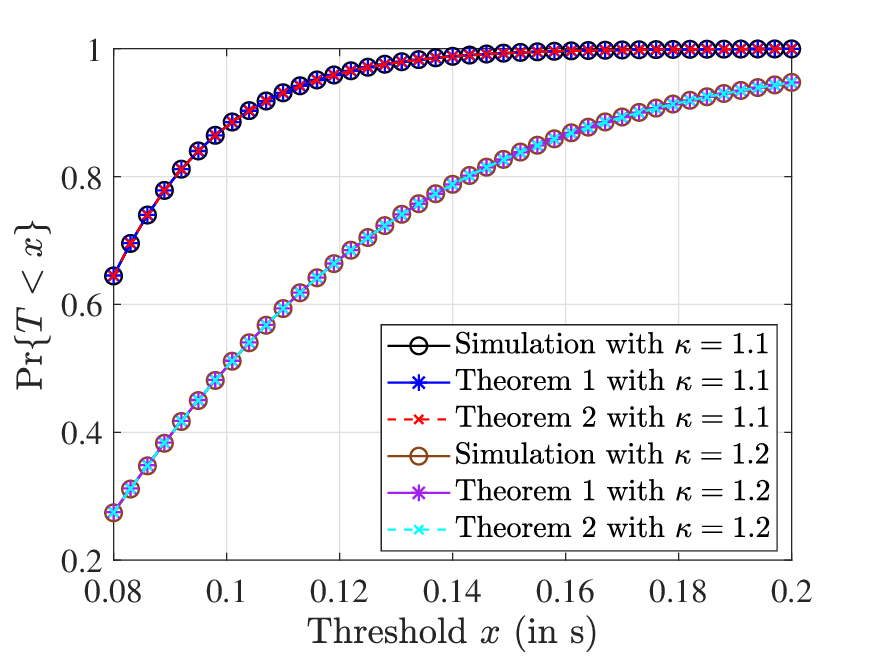}}
		\subfigure[\R{Approximation error of  $T$}]{\includegraphics[width=4.8cm]{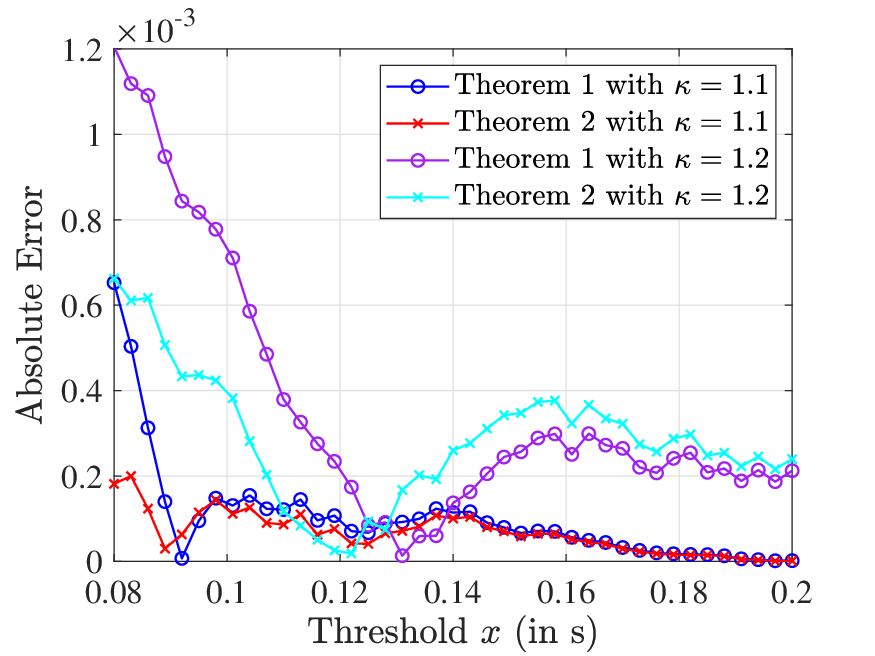}}
		\caption{$T_{\rm ET}$ and $T$ with $t_{\rm u}=5$ ms, $N_{\rm CD}^{(1.1)}=5$, $N_{\rm VI}=1$, $\epsilon_{\rm CD}=10^{-1}$, and $\epsilon_{\rm VI}=10^{-3}$.}
		\label{fig6}
	\end{minipage}
	\hspace{0.6mm}
	\begin{minipage}[b]{0.24\linewidth}
		\centering
		\subfigure[CDF of $T_{\rm ET}$.]{\includegraphics[width=4.8cm]{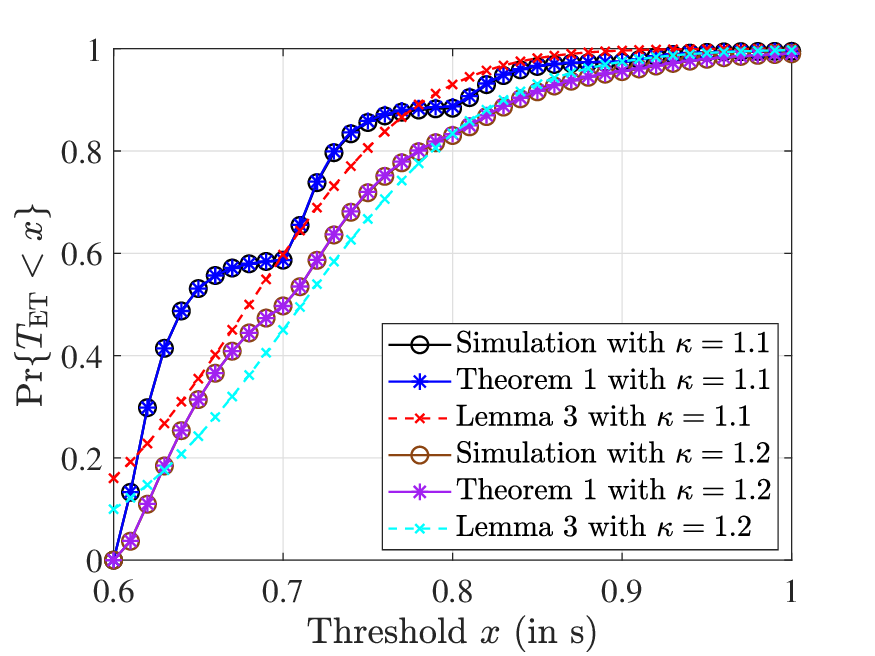}}
		\subfigure[Approximation error of $T_{\rm ET}$.]{\includegraphics[width=5cm]{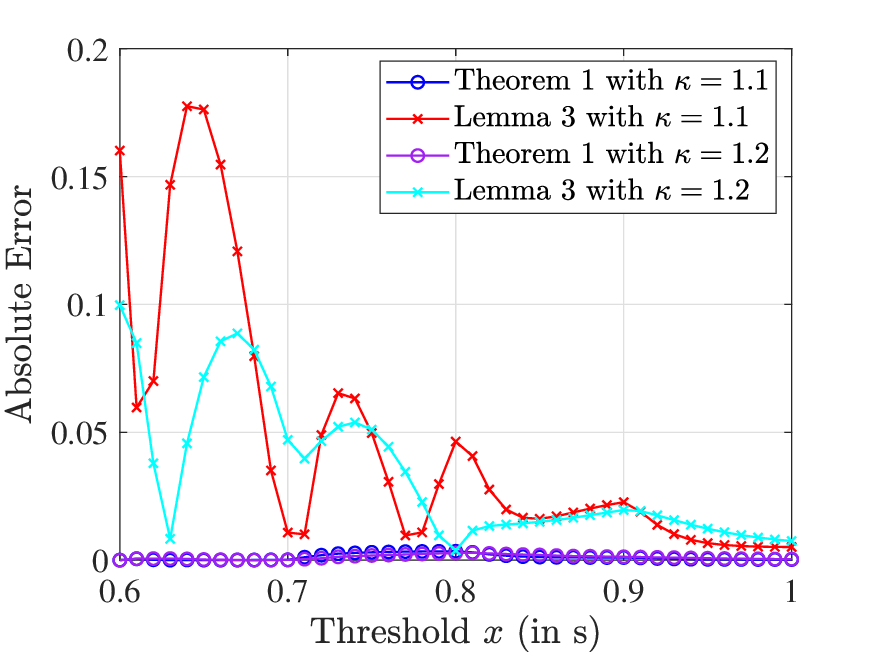}}
		\subfigure[CDF of $T$.]{\includegraphics[width=4.8cm]{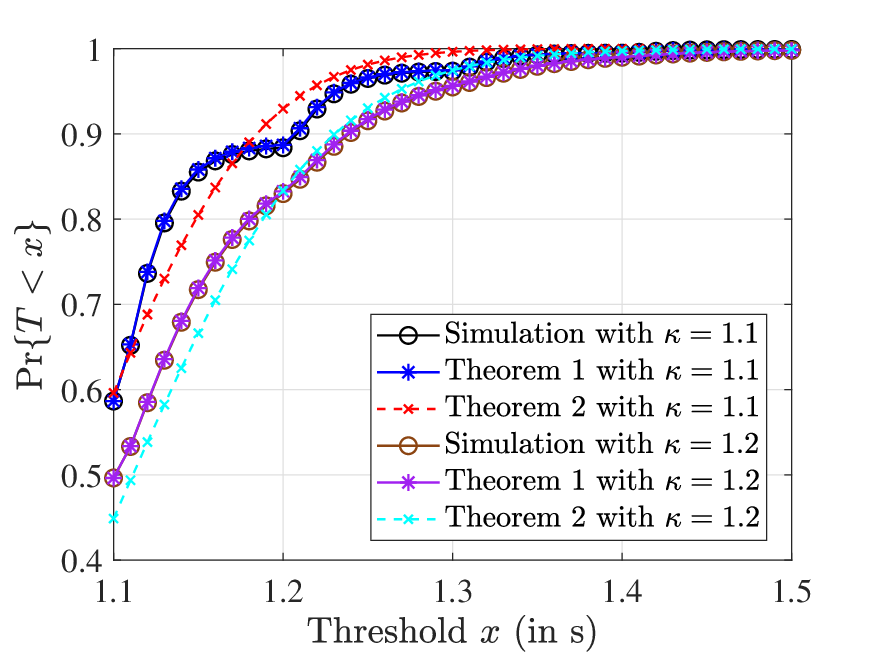}}
		\subfigure[\R{Approximation error of  $T$}]{\includegraphics[width=4.8cm]{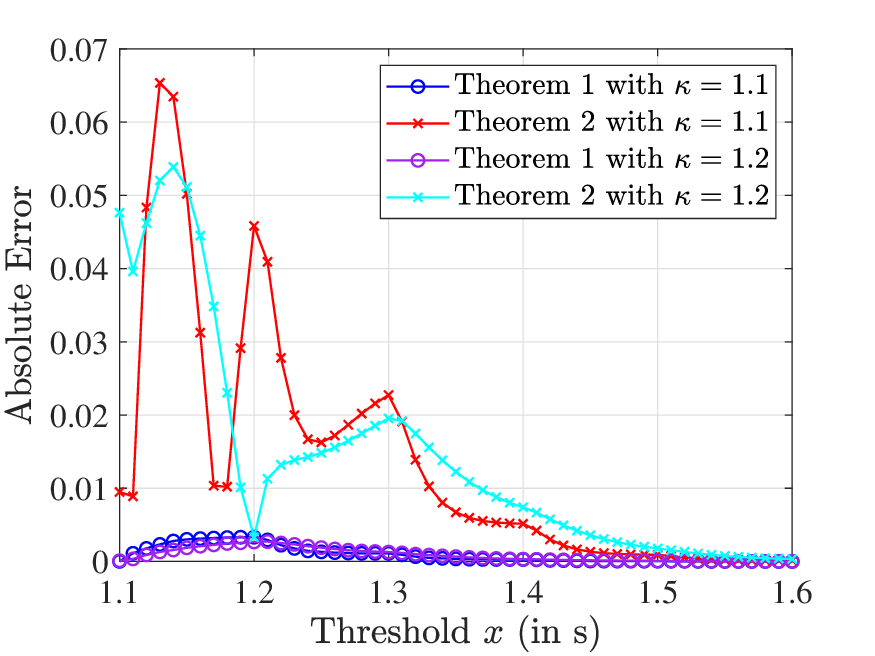}}
		\caption{$T_{\rm ET}$ and $T$ with $t_{\rm u}=100$ ms, $N_{\rm CD}^{(1.1)}=5$, $N_{\rm VI}=1$, $\epsilon_{\rm CD}=10^{-1}$, and $\epsilon_{\rm VI}=10^{-3}$.}
		\label{fig7}		
	\end{minipage}
\end{figure*}

\begin{figure*}[t]
	\begin{minipage}[t]{0.5\linewidth}
		\subfigure[CDF of $T_{\rm CL}$.]{\includegraphics[width=4.4cm]{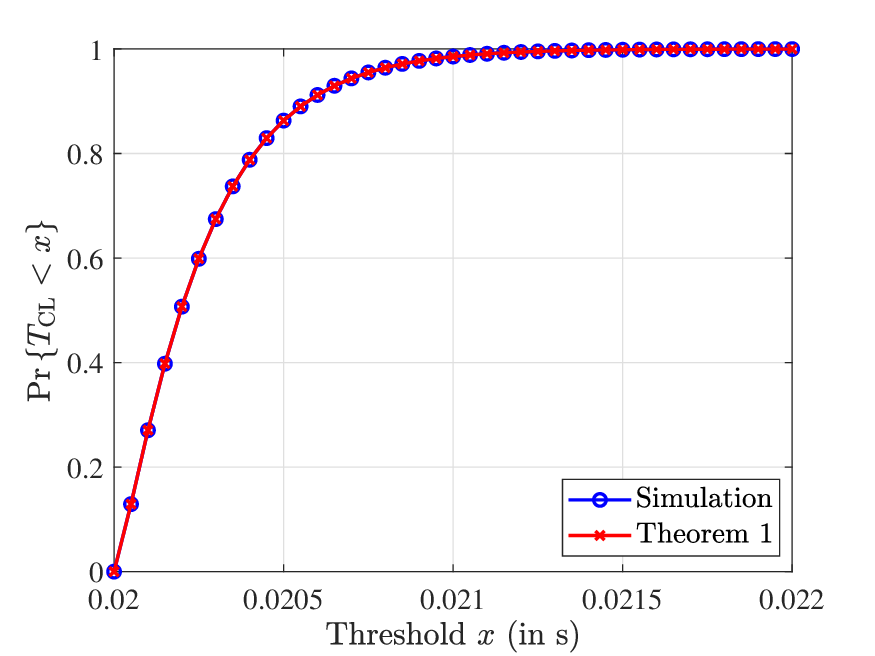}}
		\subfigure[Approximation error of  $T_{\rm CL}$.]{\includegraphics[width=4.4cm]{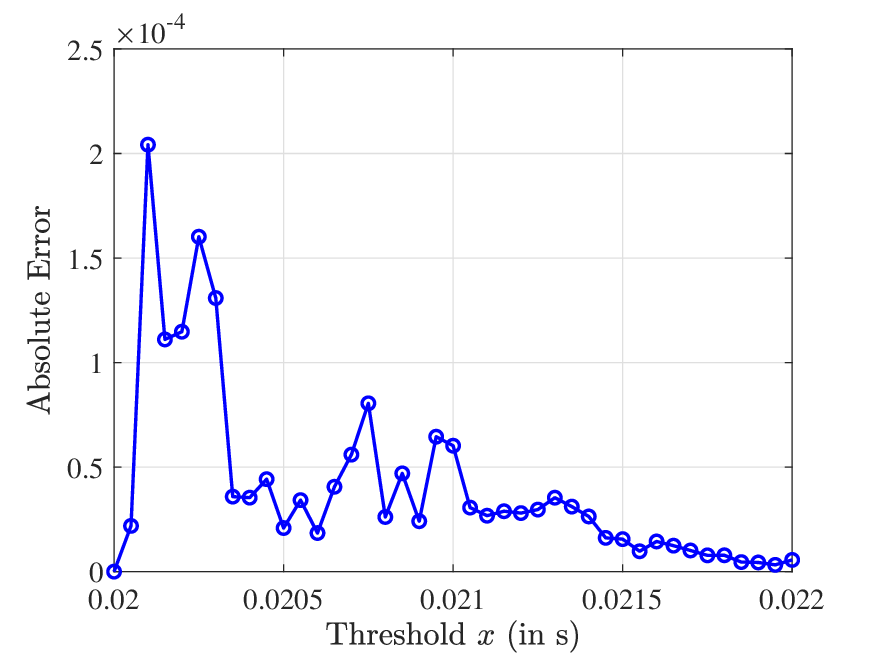}}
		\caption{$T_{\rm CL}$ with $t_{\rm u}=5$ ms.}
	\end{minipage}
\hspace{0.6mm}
	\begin{minipage}[t]{0.5\linewidth}
		\subfigure[CDF of $T_{\rm CL}$.]{\includegraphics[width=4.4cm]{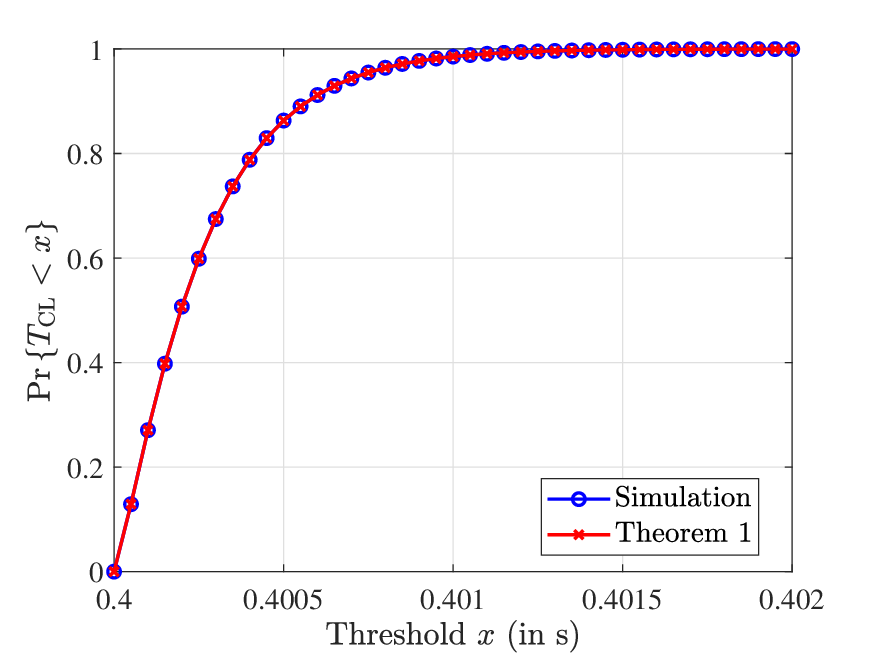}}
		\subfigure[Approximation error of  $T_{\rm CL}$.]{\includegraphics[width=4.4cm]{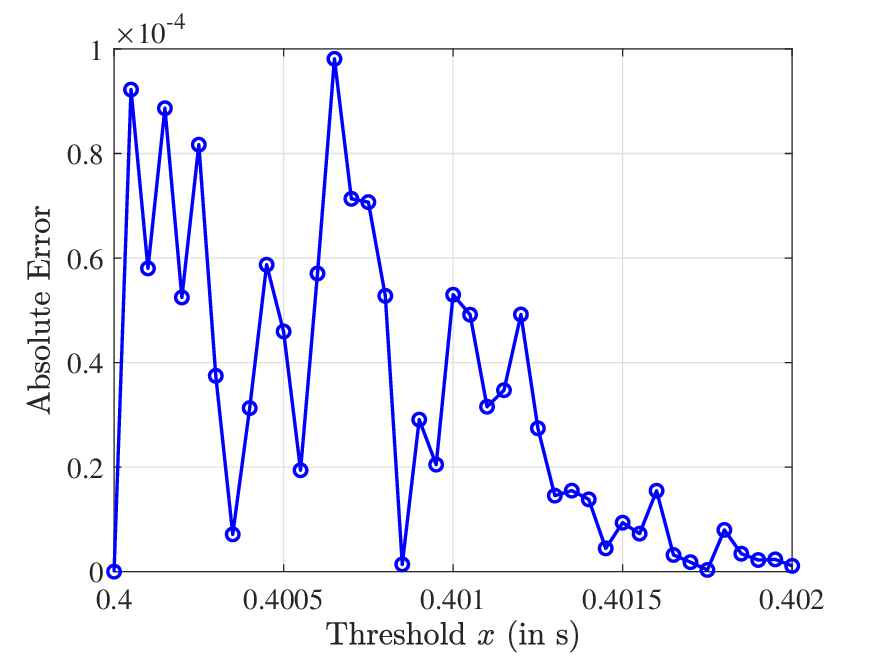}}
		\caption{$T_{\rm CL}$ with $t_{\rm u}=100$ ms.}
	\end{minipage}
\end{figure*}
	
	\R{Based on Eqs. \eqref{add6} and \eqref{add7}, we can perform optimization by using gradient-based method. It remains open to show whether $\frac{{\rm d}v_x}{{\rm d}\kappa}$ is bounded when $Z(\kappa)\to 0$. At least we can use the clipped gradient in the gradient-descent method.}

	\section{Simulation and Numerical Results}
	In this section, we validate the proposed framework through Monte Carlo simulations and numerical results.
	The parameters of the simulation and numerical calculations are mainly based on \cite{suman2023} and \cite{li2019}. Specifically, we set $\alpha_{\rm MEC}=1.25$, $\alpha_{\rm RA}=1.5$, $\beta_{\rm LL}=\beta_{\rm VI}=1$, $N_{\rm HL}=3$, $N_{\rm LL}=1$, $n_{\rm HL}=128$ KB, $\epsilon_{\rm HL}=\epsilon_{\rm LL}=10^{-5}$, $\chi_{\rm RA}=5$ GHz, $\chi_{\rm MEC}=15$ GHz, $\Delta=10^{-5}$, and $n_{\rm ET}=1$ MB. $\zeta_{\rm c}(\kappa)$ and $\zeta_{\rm d}(\kappa)$ follow Eq. \eqref{cpm1} and Eq. \eqref{decom1}, respectively, with $\psi=3.5$ and $\omega_0=0.1$.

	In Figs. \ref{fig4}-\ref{fig7}, we present the simulation and numerical results of \R{CDFs} of $T_{\rm ET}$ and $T$. \R{The simulation and numerical results of the CDF of $T_{\rm CL}$ are provided in Figs. 9 and 10.} In these figures, we let $N_{\rm CD}^{(1.1)}$ denote the number of compressed-data packets with $\kappa=1.1$. Thus, the number of compressed-data packets with $\kappa=1.2$ is given by $N_{\rm CD}^{(1.2)}=\left\lceil \frac{1.1}{1.2} N_{\rm CD}^{(1.1)}\right\rceil$. The curves with the legend `Simulation' refer to the Monte-Carlo simulation results averaged over $10^{7}$ independent simulations, while the curves with the legend `Theorem 1', `Lemma 3', and `Theorem 2' represent the numerical results of Alg. 1, Eq. \eqref{coro11}, and Eq. \eqref{th2mm}, respectively. We considered four scenarios in Figs. \ref{fig4}-\ref{fig7}. Fig. \ref{fig4} represents the general scenario. Fig. \ref{fig5} represents the scenario with a large number of packets. Fig. \ref{fig6} represents the scenarios with a small $t_{\rm u}$ and a small number of packets. Fig. \ref{fig7} represents the extreme scenario in which the number of packets is small and $t_{\rm u}$ is large. By comparing the simulation and numerical results, we find that both Theorem 1, Lemma 3, and Theorem 2 perform well in Figs. \ref{fig4}, \ref{fig5}, and \ref{fig6}. However, Theorem 1 \R{significantly outperforms} Lemma 3 in Fig. \ref{fig7}(a) and Theorem 2 in Fig. \ref{fig7}(c). 
	\R{The results in these four figures validate the accuracy of Theorem 1. When $\Pr\{T>x\}=0.01$, the absolute errors by adopting the approximation method in Theorem 1 are on the order of $10^{-5}$ in Figs. \ref{fig4}-\ref{fig6} and is on the order of $10^{-4}$ in Fig. \ref{fig7}, which are at least two orders of magnitude smaller than 0.01. Besides, as we analyzed in Section IV, Lemma 3 can be applied when $t_{\rm u}$ is small or the number of packets is large. The results in Figs. \ref{fig4}, \ref{fig5}, and \ref{fig6} show the effectiveness of Lemma 3 and Theorem 2.  The reason that Lemma 3 and Theorem 2 do not perform well in Fig. \ref{fig7} is that the discrete components of $T_{\rm ET}$ and $T$ are dominant, which is evident from the simulation curves in Fig. \ref{fig7}. Under this condition, using CLT to conduct continuous approximation leads to bad performance. In summary, Theorem 1 provides high accuracy in most closed-loop scenarios. Except for the extreme scenario where the number of packets is small or the transmission time of one packet is large, Lemma 3 and Theorem 2 offer satisfying accuracy. The high accuracy of Theorem 1 makes it capable of satisfying the specified constraint by setting the constraint stricter. It is an important future work to theoretically prove the robustness of the SPA-based method used in optimization problems.}
	
	\begin{figure*}[t]
		\begin{minipage}[b]{0.24\linewidth}
			\subfigure[Case 1.]{\includegraphics[width=4.8cm]{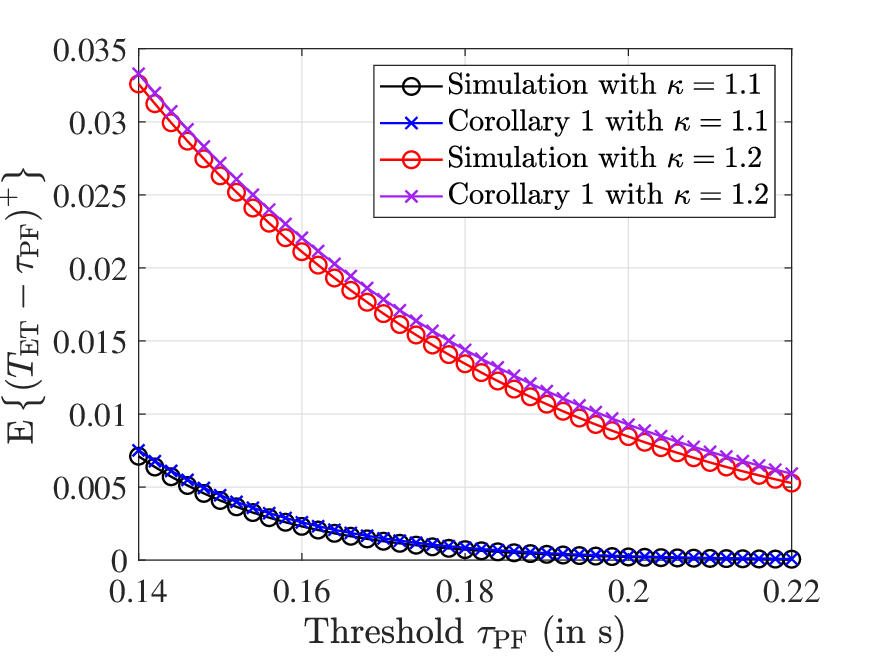}}
		\end{minipage}
		\begin{minipage}[b]{0.24\linewidth}
			\subfigure[Case 2.]{\includegraphics[width=4.8cm]{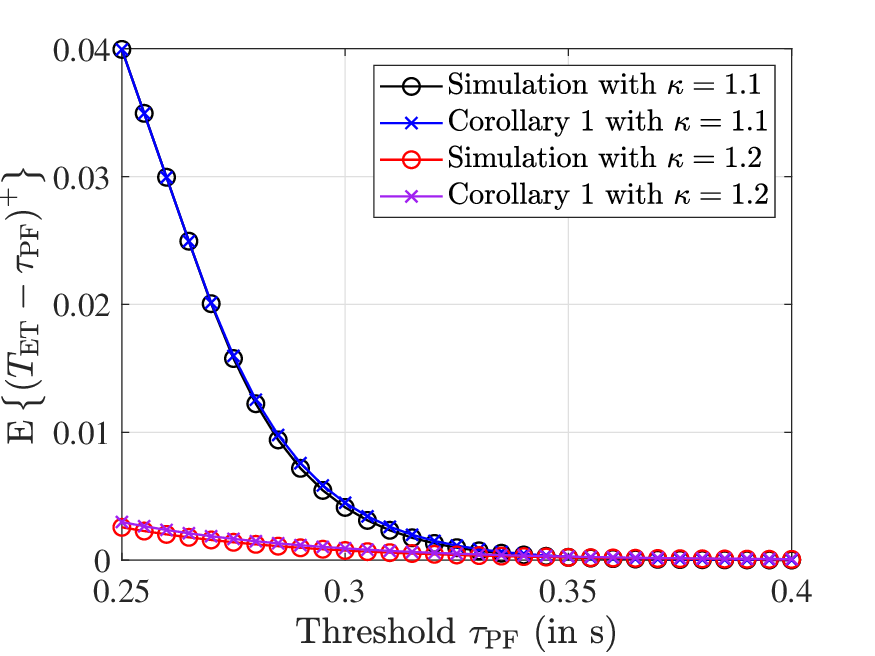}}
		\end{minipage}
		\begin{minipage}[b]{0.24\linewidth}
			\subfigure[Case 3.]{\includegraphics[width=4.8cm]{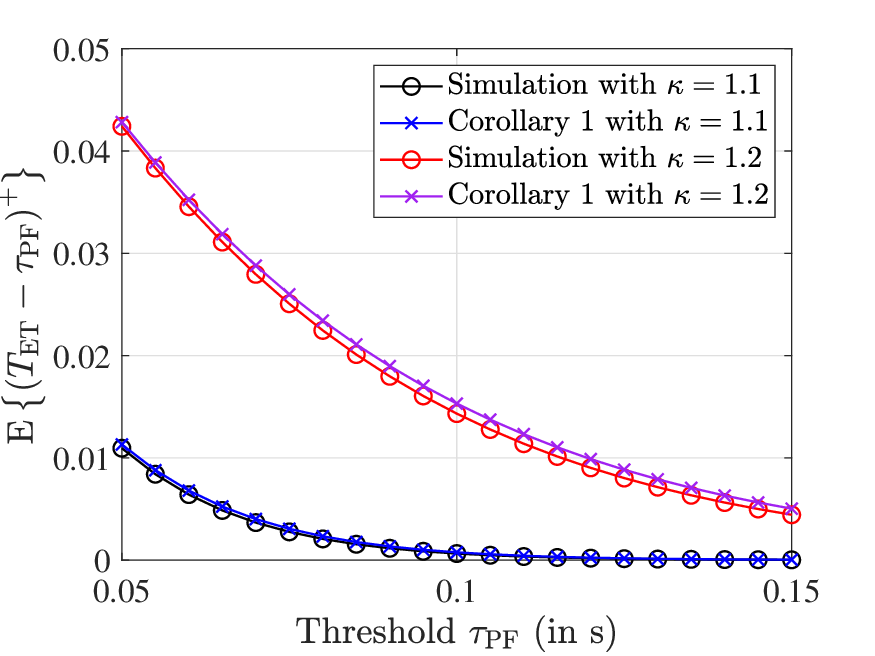}}
		\end{minipage}
		\begin{minipage}[b]{0.24\linewidth}
			\subfigure[Case 4.]{\includegraphics[width=4.8cm]{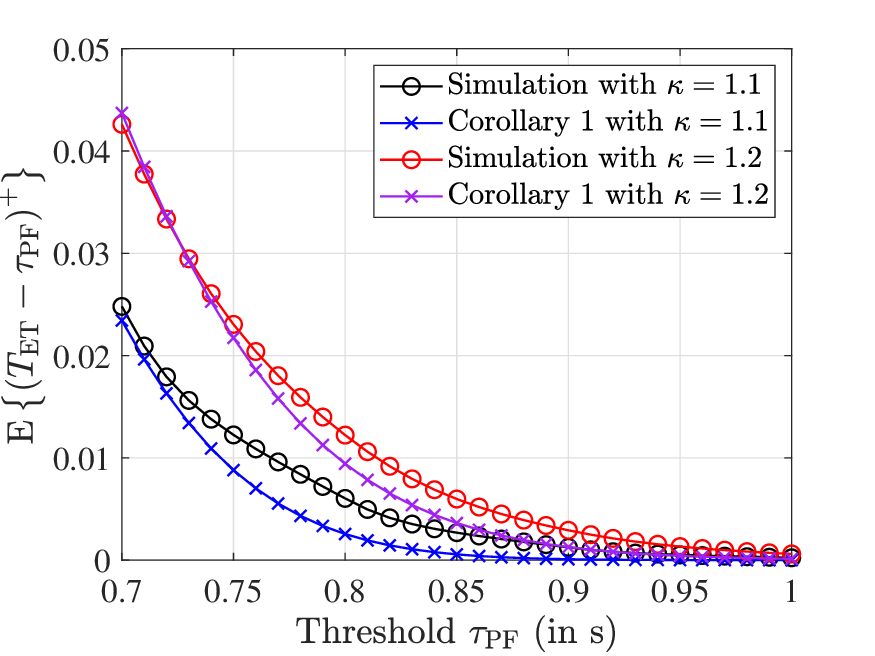}}
		\end{minipage}
		\caption{Approximation of $\mathbb{E}\left\{\left(T_{\rm ET}-\tau_{\rm PF}\right)^+\right\}$ versus $\tau_{\rm PF}$.}
		\label{fig8}
	\end{figure*}
	
	\begin{figure*}[t]
		\begin{minipage}[b]{0.245\linewidth}
			\subfigure[\R{$\varepsilon=1$.}]{\includegraphics[width=4.8cm]{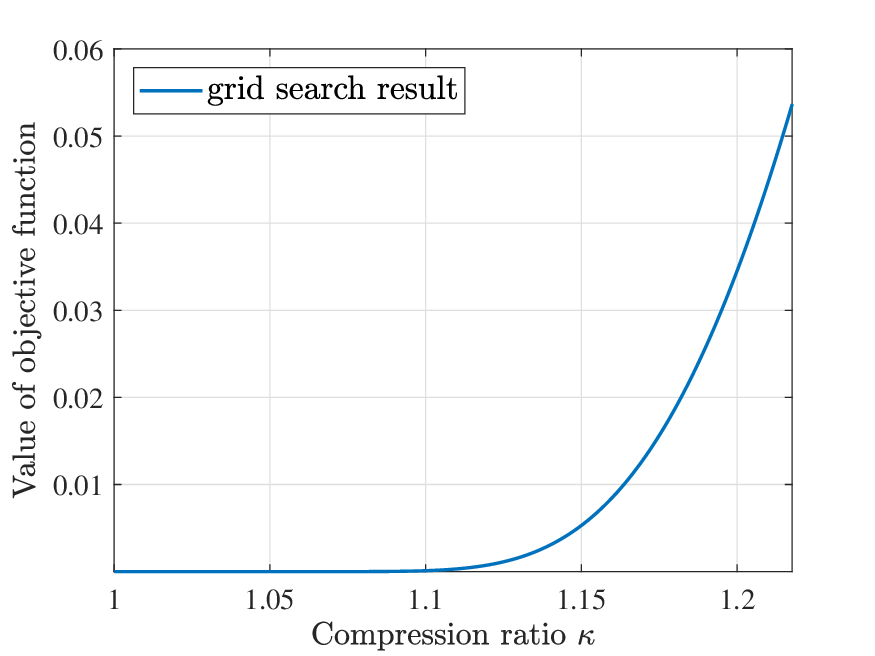}}
		\end{minipage}
		\begin{minipage}[b]{0.245\linewidth}
			\subfigure[\R{$\varepsilon=0.2$.}]{\includegraphics[width=4.8cm]{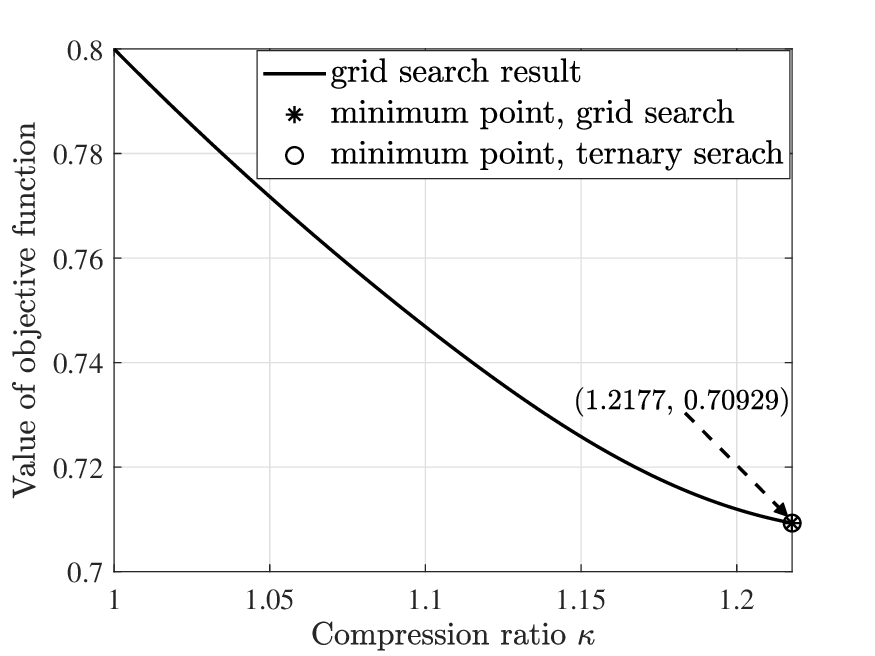}}
		\end{minipage}
		\begin{minipage}[b]{0.245\linewidth}
			\subfigure[\R{$\varepsilon=0.5$.}]{\includegraphics[width=4.8cm]{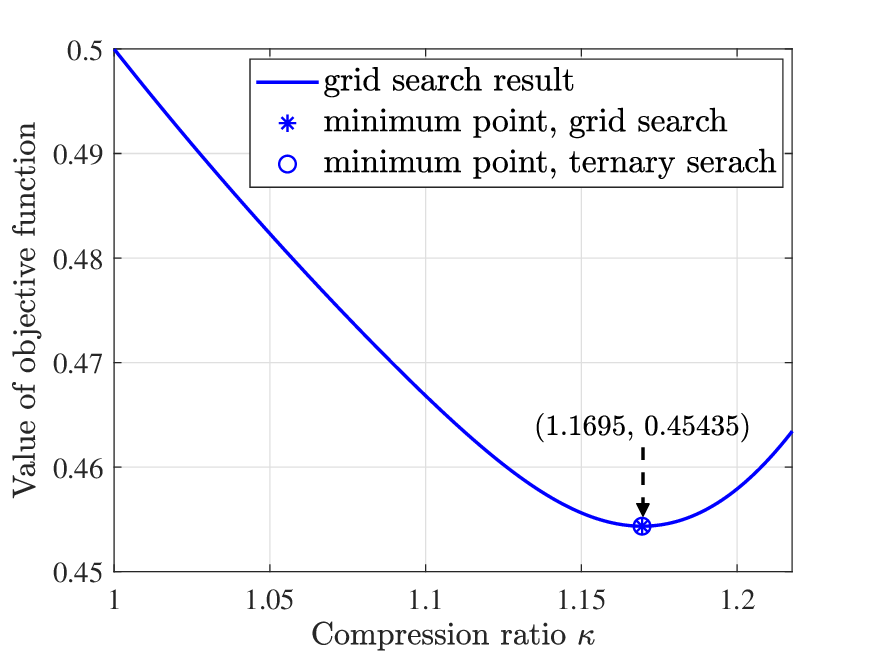}}
		\end{minipage}
		\begin{minipage}[b]{0.245\linewidth}
			\subfigure[\R{$\varepsilon=0.8$.}]{\includegraphics[width=4.8cm]{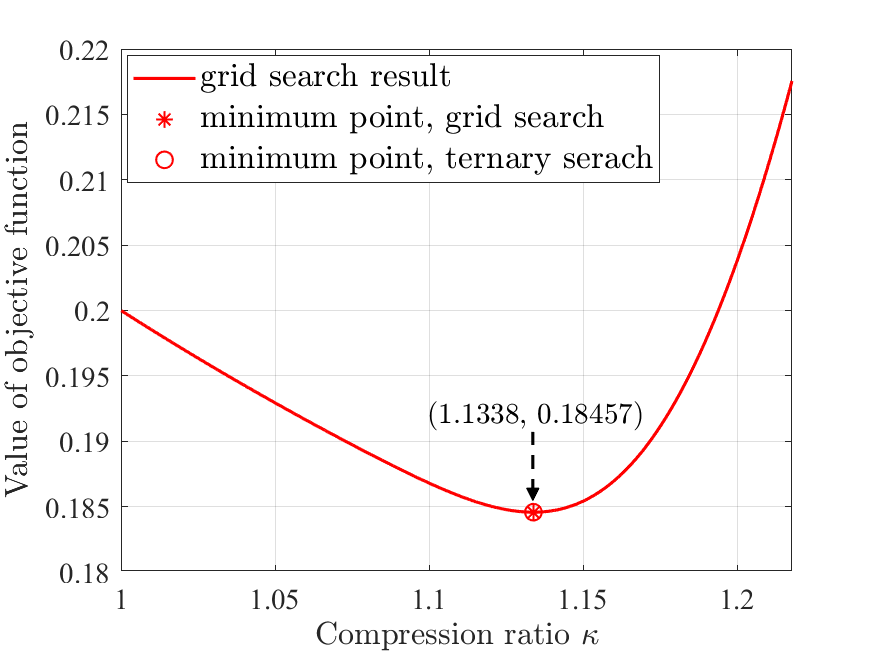}}
		\end{minipage}
		\caption{\R{Value of the objective function versus the compression ratio $\kappa$ with $\tau_{\rm PF}=0.25$ s and $T_{\rm th}=0.3$ s.}}
		\label{fig10}
	\end{figure*}

		\R{Several intriguing details emerge from our analysis.} First, we find that Theorem 2 seems to perform better than Theorem 1 in some points. \R{This phenomenon may arise from the threshold $\Delta$ set in Alg. 1, which limited the precision of the results.} Second, \R{as shown} in Fig. \ref{fig7}, the curves of Theorem 1 match well with those of the simulation results. This result shows the universality of Theorem 1 even \R{in extreme scenarios}.  \R{Moreover, we observe that the CDF does not always decrease with an increase in $\kappa$, as illustrated in Figs. \ref{fig5}(a) and \ref{fig5}(c). This behavior complicates the determination of the optimal compression ratio since the CDF with a given threshold $x$ is not monotonic with respect to $\kappa$.} However, when $x$ is large \R{(e.g., when $1-\Pr\{T_{\rm ET}<x\}$ is small)}, the CDF \R{tends to decrease as $\kappa$ increases}. Since we mainly focus on the threshold with which $1-\Pr\{T_{\rm ET}<x\}$ is small, it is important to \R{identify} and prove the condition \R{under which} $1-\Pr\{T_{\rm ET}<x\}$ decreases with increasing $\kappa$ for a given $x$ in future work.

	We then present the simulation and numerical results of $\mathbb{E}\left\{\left(T_{\rm ET}-\tau_{\rm PF}\right)^+\right\}$ in Fig. \ref{fig8}. The parameter settings in Figs.  \ref{fig8}(a)-(d) are identical with those in Figs. \ref{fig4}-\ref{fig7}. The curves with the legend `Simulation' refer to the Monte-Carlo simulation results averaged over $10^{7}$ independent simulations, while the curves with the legend `Corollary 1' represent the numerical results of Eq.  \eqref{theo21}. \R{Notably, the numerical results closely match the simulation results in Fig. \ref{fig8}(b), confirming the effectiveness of Corollary 1.} Moreover, the numerical results in Figs. \ref{fig8}(a) and \ref{fig8}(c) also \R{demonstrate} satisfactory performance. However, the difference between the numerical and simulation results becomes obvious in Fig. \ref{fig8}(d). \R{In this extreme scenario, generalization of Eq. \eqref{theo21} is required to achieve an efficient and accurate approximation for the conditional expectation, which is an important future work.}
	
\begin{figure}[t]
		\centerline{\includegraphics[width=8.5cm]{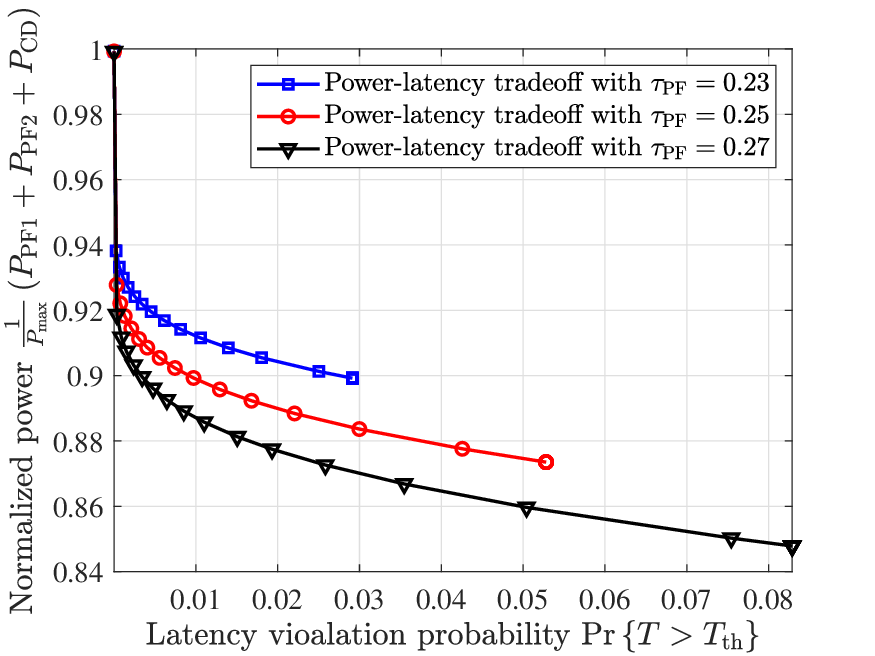}}
		\caption{\R{Power-latency tradeoff in the considered system.}}
		\label{fig11}
	\end{figure}

	\R{Finally, we present the numerical results of solving the optimization problems outlined in Section IV-D. Unless otherwise specified, the system parameters of Figs. \ref{fig10} and \ref{fig11} are the same as those in Fig. \ref{fig4}. We set $(N_0B_j)/d_j^{-\ell_j}=1$ in Figs. \ref{fig10} and \ref{fig11}. Given $\tau_{\rm PF}$, $P_{\rm max}$ is set as the sum of $P_{\rm PF1}+P_{\rm PF2}$ satisfying Eq. \eqref{ppf} and $P_{\rm CD}$ with $\kappa=1$. By this means, the power term in Problem \eqref{opj} is normalized.}
	
	\R{In Fig. \ref{fig10}, we show the value of the objective function of Problem \eqref{opj} versus the compression ratio $\kappa$. The solid curves in Fig. \ref{fig10} represent the results of grid search with granularity $10^{-4}$. We find that the objective function is unimodal with respect to $\kappa$, as shown in Fig. \ref{fig10}. Therefore, we adopt the ternary search to find the minimum points, which are marked by circles in Fig. \ref{fig10}. The minimum points obtained by grid search are marked by ``*'' in Fig.~\ref{fig10}. The results of the ternary search match well with those of the grid search, indicating that advanced algorithms may efficiently solve this problem. Besides, $\varepsilon=1$ in Fig. \ref{fig10}(a) refers to the case in which the objective function is $\Pr\{T>T_{\rm th}\}$. We see that under the given system parameters, $\Pr\{T>T_{\rm th}\}$ is a convex function of $\kappa$. Since the power term in the objective function of Problem \eqref{opj} is also a convex function of $\kappa$, the overall objective function of Problem \eqref{opj} is a convex function of $\kappa$. It is an important future work to characterize the conditions under which the optimization problem is convex. Moreover, the results in Fig. \ref{fig10} show that the optimal $\kappa$ decreases with the increase of $\varepsilon$. This is because $\Pr\{T>T_{\rm th}\}$ is an increasing function of $\kappa$ and the power term is a decreasing function of $\kappa$. With a larger $\varepsilon$, the weight of $\Pr\{T>T_{\rm th}\}$ becomes higher, leading to the decrease of $\kappa$. } 
	
	\R{From the analysis of Fig. \ref{fig10}, we see that there exists an optimal value of $\kappa$, highlighting a tradeoff between the performance of latency and power in the considered system. In Fig. \ref{fig11}, we present the power-latency tradeoff with different $\tau_{\rm PF}$. Note that the feasible interval of $\kappa$ is influenced by the value of $\tau_{\rm PF}$. Thus, the tradeoff curves have varying ranges as shown in Fig. \ref{fig11}. From Fig. \ref{fig11}, we find that the optimal power-latency tradeoff is convex in the considered system. 	 As shown in Fig. \ref{fig11}, the power consumption is higher with a smaller $\tau_{\rm PF}$ when achieving the same latency violation probability. This is because a smaller $\tau_{\rm PF}$ refers to a larger $P_{\rm PF1}+P_{\rm PF2}$ according to Eq. \eqref{ppf}. Besides, a smaller $\tau_{\rm PF}$ imposes stricter requirements on the ET latency, which results in a larger $P_{\rm CD}$. These results may provide instructions for the choices of system parameters.}

	\section{Conclusion}

	In this paper, we focused on formulating a unified timing-analysis framework and proposing a general method to characterize the distribution of the closed-loop timing performance in CGC systems. The multi-modal feedback was divided into PF and ET data due to the heterogeneity of the feedback mechanism. Covering timeliness, jitter, and reliability, we developed a unified framework of timing analysis for CGC systems with various configurations, which includes both average-based and tail-based constraints. The randomness of computation, compression, and communication were incorporated into the closed-loop timing analysis. To address the difficulties resulting from the sum of multiple random latency components, an SPA-based method was conceived, which helped us derive the analytical expression of the closed-loop latency and other timing metrics. Important future topics include further optimization based on this unified framework, the introduction of emerging timing metrics into this framework, and the generalization of the approximation for the conditional expectation of timing metrics.

	\appendices
	
	\section{Proof of Lemma 1}
	\R{The essence of the proof is to use the LR formula \cite[Eq. (1.21)]{LR}. For a detailed derivation of the LR formula, please refer to \cite[Ch. 2]{LR}. Here we give the sketch of how to use Eq. (1.21) of \cite{LR}.} We first obtain the cumulant generating function (CGF) of $T^{\rm c}_{\rm ET}+T^{\rm d}_{\rm ET}+T^{\rm cp}_{\rm VI}$.  The CGF of gamma random variable $X_1\sim \text{Gamma}(\alpha,\beta)$ is given by
	\begin{equation}\label{cgfgamma}
		K_{X_1}(s)=-\alpha\ln\left(1-\frac{s}{\beta}\right), \quad s<\beta.	
	\end{equation}

	Based on Eq. \eqref{cgfgamma}, we obtain the CGF of $T^{\rm c}_{\rm ET}+T^{\rm d}_{\rm ET}+T^{\rm cp}_{\rm VI}$, which is given in Eq. \eqref{cgftmd}, where $s<\min\left\{ \frac{\chi_{\rm RA}\alpha_{\rm RA}}{n_{\rm ET}\zeta_{\rm c}(\kappa)}, \frac{\chi_{\rm MEC}\alpha_{\rm MEC}\kappa}{n_{\rm ET}\zeta_{\rm d}(\kappa)},   \frac{\chi_{\rm MEC}\beta_{\rm VI}}{n_{\rm ET}}   \right\}$. To obtain the CDF of $T^{\rm c}_{\rm ET}+T^{\rm d}_{\rm ET}+T^{\rm cp}_{\rm VI}$, we need to solve the saddlepoint equation to obtain the saddlepoint $s_x$. The saddlepoint equation is given by
	$L'(s_x)=x$,
	where $L'(s)$ is given by 
	\begin{equation}\label{spaequsub}
		\begin{aligned}
			\!\! \! L'(s)\!=\!\frac{\alpha_{\rm RA}}{\frac{\chi_{\rm RA}\alpha_{\rm RA}}{n_{\rm ET}\zeta_{\rm c}(\kappa)}-s}\!+\!\frac{\alpha_{\rm MEC}}{\frac{\chi_{\rm MEC}\alpha_{\rm MEC}\kappa}{n_{\rm ET}\zeta_{\rm d}(\kappa)} -s }
			\!+\!\frac{\alpha_{\rm MEC}}{\frac{\chi_{\rm MEC}\beta_{\rm VI}}{n_{\rm ET}}-s}.
		\end{aligned}
	\end{equation}
	
	$L'(s_x)$ is an increasing function of $s_x$. Thus, the inverse function of $L':s_x\mapsto x$ exists,  which is denoted by $\Xi_1:x\mapsto s_x$. There is a unique solution of the saddlepoint equation, i.e., $\Xi_1(x)$, which can be solved by binary search.

	Then, we divide the following proof into two cases. Let $\theta=\mathbb{E}\{T^{\rm c}_{\rm ET}+T^{\rm d}_{\rm ET}+T^{\rm cp}_{\rm VI}\}$ as shown in \eqref{expectmdc}. The first case is that $x\neq \theta$ holds. \R{According to the LR formula \cite[Eq. (1.21)]{LR}}, we obtain 
	\begin{equation}
		F_1(x)=\Phi(v_x)+\phi(v_x)\left(\frac{1}{v_x}-\frac{1}{u_x}\right),
	\end{equation}
	where $v_x={\rm sign}(s_x)\sqrt{2(xs_x-L(s_x))}$, and $u_x=s_x\sqrt{L''(s_x)}$. The second-order derivative of $L(s)$ is given by
	\begin{equation}\label{2orderderi}
		\begin{aligned}
			L''(s)=\frac{\alpha_{\rm RA}}{\left(\frac{\chi_{\rm RA}\alpha_{\rm RA}}{n_{\rm ET}\zeta_{\rm c}(\kappa)}-s\right)^2}&+\frac{\alpha_{\rm MEC}}{\left(\frac{\chi_{\rm MEC}\alpha_{\rm MEC}\kappa}{n_{\rm ET}\zeta_{\rm d}(\kappa)} -s \right)^2}\\
			&+\frac{\alpha_{\rm MEC}}{\left(\frac{\chi_{\rm MEC}\beta_{\rm VI}}{n_{\rm ET}}-s\right)^2}.
		\end{aligned}
	\end{equation}

	Thus, by substituting Eqs. \eqref{cgftmd} and \eqref{2orderderi} into the expressions of $v_x$ and $u_x$, we obtain the first expression in Eq. \eqref{lemma11}.

	The second case is that $x=\theta$ holds. \R{According to the LR formula \cite[Eq. (1.21)]{LR}, $\Pr\{T_{\rm ET}<\theta\}$ is given by}
	\begin{equation}\label{ex0}
		\begin{aligned}
			F_1(x)&=\frac{1}{2}+\frac{L'''(0)}{6\sqrt{2\pi }L''(0)^{3/2}}.
		\end{aligned}
	\end{equation}
	We let $\iota_1=L''(0)^{3/2}$ and $\iota_2=L'''(0)$, which are given in Eqs. \eqref{const} and \eqref{const1}, respectively. 	By substituting Eqs. \eqref{const} and \eqref{const1} into Eq. \eqref{ex0}, we obtain Eq. \eqref{lemma11}. 
	
	\R{For the PDF, we adopt the saddlepoint density functions in [Eqs. (1.4) and (1.5)]\cite{LR}. For a detailed derivation of saddlepoint density functions, please refer to \cite[Ch. 2]{LR}. By substituting Eqs. \eqref{cgftmd} and \eqref{2orderderi} into the Eqs. (1.4) and (1.5) of \cite{LR}, we obtain Eq. \eqref{lemma13}.}

	\section{Proof of Lemma 2}
	Similar to the proof of Lemma 1, we shall first obtain the CGF of  $T^{\rm cm}_{\rm CD}+T^{\rm cm}_{\rm VI}$. The CGF of negative binomial random variable $X_2\sim \text{NB}(N,p)$ is given by
	\begin{equation}\label{cgfnb}
		K_{X_2}(s)=N\left(s+\ln p-\ln\left(1-(1-p)e^s\right)\right),
	\end{equation}
	where $s<-\ln(1-p)$. Based on Eq. \eqref{cgfnb}, we obtain the CGF of $T^{\rm cm}_{\rm CD}+T^{\rm cm}_{\rm VI}$, which is given by Eq. \eqref{cgfdis1}. The first-order derivative of $J(s)$ is thus given by
	\begin{equation}\label{cgfdis2}
		\begin{aligned}
			J'(s)=\frac{N_{\rm CD}}{1-\epsilon_{\rm CD}e^{s}}+\frac{N_{\rm VI}}{1-\epsilon_{\rm VI}e^{s}},
		\end{aligned} 
	\end{equation}
	where $s<\min \left\{-\ln\epsilon_{\rm CD},-\ln\epsilon_{\rm VI} \right\}$. 
	The saddlepoint equation is thus given by $J'(s_k)=k$.	$J'(s_k)$ is an increasing function of $s_k$. Thus, the inverse function of $J':s_k \mapsto x$ exists,  which is denoted by $\Xi_2:x\mapsto s_k$. The solution of the saddlepoint equation is thus given by $\Xi_2(k)$.
	
	Based on Eq. \eqref{cgfdis2}, we can also obtain
	\begin{equation} \label{cgfdis3}
		\begin{aligned}
			J''(s)=\frac{\epsilon_{\rm CD}N_{\rm CD}e^{s}}{\left(1-\epsilon_{\rm CD}e^{s}\right)^2}+\frac{\epsilon_{\rm VI}N_{\rm VI}e^{s}}{\left(1-\epsilon_{\rm VI}e^{s}\right)^2},
		\end{aligned}
	\end{equation}
	where $s<\min \left\{-\ln\epsilon_{\rm CD},-\ln\epsilon_{\rm VI} \right\}.$ 	\R{According to the saddlepoint mass function \cite[Eq. (1.13)]{LR}}, we obtain the PMF for $k>N_{\rm CD}+N_{\rm VI}$, which is the interior of the span of the support of $T^{\rm cm}_{\rm CD}+T^{\rm cm}_{\rm VI}$. For $k=N_{\rm CD}+N_{\rm VI}$, the probability can be simply obtained as $(1-\epsilon_{\rm CD})^{N_{\rm CD}}(1-\epsilon_{\rm VI})^{N_{\rm VI}}$.

	\section{Proof of Lemma 4}
		\R{By rewriting Eq. \eqref{dcc}, we obtain that}
	\begin{equation}
		\R{P_j=\varphi_j^{-c_3-1}\left(c_2+(\tau_j)^{-c_3}\right).}
	\end{equation}
	
	\R{$P_j$ is a convex function of $\tau_j$ since $c_3>0$ and $\varphi_j>0$. Thus, the conclusion in Lemma 4 is obtained by solving the following convex optimization problem:}
\begin{subequations}
	\R{\begin{align}
					\min_{\tau_{\rm PF1},\tau_{\rm PF2}}\quad & \sum_{j\in\{{\rm PF1},{\rm PF2}\}}\varphi_{j}^{-c_3-1}\left(c_2+(\tau_j)^{-c_3}\right) \\
			\mbox{s.t.}\quad  & \tau_{\rm PF1}+\tau_{\rm PF2}\leq \tau_{\rm PF}.
		\end{align}}
	\end{subequations}
\R{With the help of Karush–Kuhn–Tucker (KKT) conditions, the optimal solutions follows}
\begin{equation}\label{kkt}
\R{	\begin{cases}
		\lambda(\tau_{\rm PF1}+\tau_{\rm PF2}-\tau_{\rm PF})=0, \\
		-c_3\varphi_j^{-c_3-1}\tau_j^{-c_3-1}+\lambda=0, \quad j\in\{{\rm PF1}, {\rm PF2}\},
	\end{cases}}
\end{equation}
\R{where $\lambda>0$ is the Lagrange multiplier. By solving equations \eqref{kkt}, we obtain $\tau_{\rm PF1}=\frac{\tau_{\rm PF}\varphi_{\rm PF2}}{\varphi_{\rm PF1}+\varphi_{\rm PF2}}$ and $\tau_{\rm PF2}=\frac{\tau_{\rm PF}\varphi_{\rm PF1}}{\varphi_{\rm PF1}+\varphi_{\rm PF2}}$. The proof is completed.}

	\section{Proof of Lemma 5}
	Let $G(\kappa)=\frac{\kappa}{\zeta_{\rm d}({\kappa})}$. For $\zeta_{\rm d}(\kappa)$ shown in Eq. \eqref{decom1}, we obtain
	\begin{equation}
		\frac{{\rm d}G(\kappa)}{{\rm d}\kappa}=\frac{e^{\psi\kappa}(1-\psi\kappa)-e^{\psi}}{\omega_0\left(e^{\psi\kappa}-e^{\psi}\right)^2}.	 
	\end{equation}
	Since $\kappa\geq 1$, $\frac{{\rm d}G(\kappa)}{{\rm d}\kappa}\leq 0$ holds with $\psi\geq1$. For $\zeta_{\rm d}(\kappa)$ shown in Eq. \eqref{decom2}, we have
	\begin{equation}
		\frac{{\rm d}G(\kappa)}{{\rm d}\kappa}=\frac{\omega_5\omega_6(1-\omega_7)\kappa^{\omega_7}+\omega_5\omega_8}{\left(\omega_5(\omega_6 \kappa ^{\omega_7}+\omega_8)\right)^2}.
	\end{equation}
	From this equation, we can find that $\omega_5\omega_6(1-\omega_7)\kappa^{\omega_7}$ is decreasing with $\kappa$ when $\omega_7>1$. Thus, the maximum value of $\omega_5\omega_6(1-\omega_7)\kappa^{\omega_7}$ is achieved at $\kappa=1$ since $\kappa\geq 1$, which is $\omega_5\omega_6(1-\omega_7)$. Thus, if $\omega_5\omega_6(1-\omega_7)+\omega_5\omega_8<0$, i.e., $\omega_7>1+\frac{\omega_8}{\omega_6}$, the maximum value of $\frac{{\rm d}G(\kappa)}{{\rm d}\kappa}<0$. 
	
	For $\omega_7\leq 1$, $\omega_5\omega_6(1-\omega_7)\kappa^{\omega_7}\geq 0$. Thus, $\frac{{\rm d}G(\kappa)}{{\rm d}\kappa}>0$.

	\begin{spacing}{1}
		\bibliographystyle{IEEEtran}
		\bibliography{mybib}
	\end{spacing}

\end{document}